\documentclass[a4,12pt]{article}
\usepackage{amstext,amsmath,amssymb}
\usepackage[margin=1in]{geometry}
\usepackage{graphicx}
\usepackage{algorithm}
\usepackage{algpseudocode}
\usepackage{fmtcount}
\usepackage[pagewise]{lineno}
\usepackage[dvipsnames,usenames]{color}
\usepackage[normalem]{ulem}
\usepackage{tabularx}
\usepackage{rotating}
\usepackage{enumerate}
\usepackage{graphicx}

\usepackage{caption}
\usepackage{subcaption}
\usepackage[maxfloats=40]{morefloats}

\usepackage[resetlabels,labeled]{multibib}
\newcites{sec}{Referenceswss}

\newtheorem{lemma}{Lemma}
\newtheorem{thm}{Theorem}

\numberwithin{equation}{section}
\numberwithin{lemma}{section}
\numberwithin{thm}{section}

\newcommand{\cur}{\mbox{\scriptsize\,current}}
\newcommand{\old}{\mbox{\scriptsize\,old}}
\newcommand{\con}{\mbox{\scriptsize con}}

\title{A convex framework for high-dimensional sparse Cholesky based covariance  
estimation}
\date{}
\author{Kshitij Khare, Sang Oh, Syed Rahman and Bala Rajaratnam}

\begin{document}
\maketitle

\begin{abstract}
Covariance estimation for high-dimensional datasets is a fundamental
problem in modern day statistics with numerous applications. In these
high dimensional datasets, the number of variables $p$ is typically
larger than the sample size $n$. A popular
way of tackling this challenge is to induce sparsity
in the covariance matrix, its inverse or a relevant transformation.
In particular, methods inducing sparsity in the Cholesky parameter of
the inverse covariance matrix can be useful as they are guaranteed to
give a positive definite estimate of the covariance matrix. Also, the estimated sparsity
pattern corresponds to a Directed Acyclic Graph (DAG) model for
Gaussian data. In recent years, two useful penalized likelihood methods for
sparse estimation of this Cholesky parameter (with no restrictions on
the sparsity pattern) have been developed. However, these methods
either consider a non-convex optimization problem which can lead to
convergence issues and singular estimates of the covariance matrix 
when $p > n$, or achieve a convex formulation by
placing a strict constraint on the conditional variance parameters. In
this paper, we propose a new penalized likelihood method for sparse
estimation of the inverse covariance Cholesky parameter that aims to
overcome some of the shortcomings of current methods, but retains
their respective strengths. We obtain a jointly convex formulation for
our objective function, which leads to convergence guarantees, even
when $p > n$. The approach always leads to a positive definite and
symmetric estimator of the covariance matrix. We establish
high-dimensional estimation and graph selection consistency, and
also demonstrate finite sample performance on simulated/real data. 
\end{abstract}

\section{Introduction} \label{sec:intro}

\noindent
In modern day statistics, datasets where the number of variables is much higher than
the number of samples are more pervasive than they have ever been. One of the 
major challenges in this setting is to formulate models and develop inferential 
procedures to understand the complex relationships and multivariate dependencies 
present in these datasets. The covariance matrix is the most fundamental 
object that quantifies relationships between the variables in multivariate datasets. 
Hence, estimation of the covariance matrix is crucial in high-dimensional problems and 
enables the detection of the most important relationships.

In particular, suppose we have i.i.d. observations ${\bf Y}_1, {\bf Y}_2, \cdots, 
{\bf Y}_n$ from a $p$-variate normal distribution with mean vector ${\bf 0}$ and 
covariance matrix $\Sigma$. Note that $\Sigma \in \mathbb{P}_p^+$, the space of 
positive definite matrices of dimension $p$. In many modern applications, the number 
of observations $n$ is much less than the number of variables $p$. In such 
situations, parsimonious models which restrict $\Sigma$ to a lower dimensional 
subspace of $\mathbb{P}_p^+$ are required for meaningful statistical estimation. Let 
$\Sigma^{-1} = T^t D^{-1} T$ denote the modified Cholesky decomposition of 
$\Omega = \Sigma^{-1}$. Here $T$ is a lower triangular matrix with diagonal entries 
equal to $1$ (we will refer to $T$ as the Cholesky parameter), and $D$ is a diagonal 
matrix with positive diagonal entries. The entries of $T$ and $D$ have a very natural 
interpretation. In particular, the (nonredundant) entries in each row of $T$ are 
precisely the regression coefficients of the corresponding variable on the preceding 
variables. Similarly, each diagonal entry of $D$ is the residual variance of the 
corresponding variable after regression on the preceding variables. 

Owing to these interpretations, various authors in the literature have considered 
sparse estimation of $T$ as a means of inducing parsimony in high-dimensional 
situations. Smith and Kohn \cite{Smith:Kohn:2002} develop a hierarchical Bayesian 
approach which allows for sparsity in the Cholesky parameter. 
Wu and Pourahmadi \cite{Wu:Pourahmadi:2003} develop a non-parametric smoothing 
approach which provides a sparse estimate of the Cholesky parameter, with a banded 
sparsity pattern. Huang et al. \cite{HLPL:2006} introduce a penalized 
likelihood method to find a regularized estimate of $\Omega$ with a sparse Cholesky 
parameter. Levina et al. \cite{LRZ:2008} develop a penalized likelihood approach 
using the so-called nested lasso penalty to provide a sparse banded estimator for the 
Cholesky parameter. Rothman et al. \cite{RLZ:2010} develop penalized 
likelihood approaches for the related but different problem of sparse estimation of 
$T^{-1}$. Shajoie and Michalidis \cite{Shajoie:Michalidis:2010} motivate 
sparsity in the Cholesky parameter $T$ as a way of estimating the skeleton graph for 
a Gaussian Directed Acyclic Graph (DAG) model. In recent parallel work, 
Yu and Bien \cite{YuBien:2010} develop a penalized likelihood approach to obtain a
tapered/banded estimator of $T$ (with possibly different bandwiths for
each row). To the best of our knowledge, the 
methods in \cite{HLPL:2006} and \cite{Shajoie:Michalidis:2010} are the only (non-
Bayesian) methods which induce a general or unrestricted sparsity pattern in the 
inverse covariance Cholesky parameter $T$. Although both these methods are quite 
useful, they suffer from some drawbacks which we will discuss below. 

Huang et al. \cite{HLPL:2006} obtain a sparse estimate of $T$ by minimizing the 
objective function 
\begin{equation} \label{eq1}
Q_{Chol} (T,D) = tr \left( T^t D^{-1} T S \right) + \log |D| + \lambda \sum_{1 \leq i < j 
\leq p} |T_{ij}|. 
\end{equation}

\noindent
with respect to $T$ and $D$, where $S = \frac{1}{n} \sum_{i=1}^n {\bf Y}_i {\bf Y}_i^T$ 
is the sample covariance matrix (note that ${\bf Y}_i's$ have mean zero). Let 
${\boldsymbol \phi}^i := (T_{ij})_{j=1}^{i-1}$ and $S_{\cdot i} := (S_{ij})_{j=1}^{i-1}$ respectively 
denote the vector of lower triangular entries in the $i^{th}$ row of $T$ and $S$ for $i = 2, 
3, \cdots, p$. Let $S_i$ denote the $i \times i$ submatrix of $S$ starting from the first row 
(column) to the $i^{th}$ row (column), for $i = 1,2, \cdots, p$. It can be established after 
some simplification (see \cite{HLPL:2006}) that 
$$
Q_{Chol} (T,D) = \left\{ \frac{S_{11}}{D_{11}} + \log D_{11} 
\right\} + \sum_{i=2}^{p} \left\{ \frac{({\boldsymbol \phi}^i)^t S_{i-1} {\boldsymbol 
\phi}^i + 2 ({\boldsymbol \phi}^i)^t S_{\cdot i} + S_{ii}}{D_{ii}} + \log D_{ii} + \lambda 
\|{\boldsymbol \phi}^i\|_1 \right\}, 
$$

\noindent
where $\|{\bf x}\|_1$ denotes the sum of absolute values of the entries of a vector ${\bf x}$. 
It follows that minimizing $Q_{Chol} (L,D)$ with respect to $L$ and $D$, is equivalent 
to minimizing 
\begin{equation} \label{eqrep}
Q_{Chol,i} ({\boldsymbol \phi}^i, D_{ii}) = \frac{({\boldsymbol \phi}^i)^t S_{i-1} {\boldsymbol 
\phi}_i + 2 ({\boldsymbol \phi}^i)^t S_{\cdot i} + S_{ii}}{D_{ii}} + \log D_{ii} + \lambda 
\|{\boldsymbol \phi}^i\|_1 
\end{equation}

\noindent
with respect to $({\boldsymbol \phi}^i, D_{ii})$ for $i = 2, 3, \cdots, p$, and setting 
$D_{11} = S_{11}$. Huang et al. \cite{HLPL:2006} propose minimizing $Q_{Chol, i}$ using 
cyclic block coordinatewise minimization, where each iteration consists of 
minimizing $Q_{Chol,i}$ with respect to ${\boldsymbol \phi}$ (fixing $D_{ii}$ at its 
current value), and then with respect to $D_{ii}$ (fixing ${\boldsymbol \phi}^i$ at 
its current value). However, this regularization approach based on minimizing 
$Q_{Chol}$ encounters a problem when $n < p$. In particular, the following lemma 
(proof provided in the supplemental document) holds. 
\begin{lemma} \label{leminfimum}
The function $Q_{Chol,i} ({\boldsymbol \phi}^i, D_{ii})$ is not jointly convex or 
bi-convex for $1 \leq i \leq p$. Moreover, if $n < p$, then 
$$
\inf_{{\boldsymbol \phi}_{n+1} \in \mathbb{R}^{n}, D_{n+1,n+1} > 0} Q_{Chol, n+1} 
({\boldsymbol \phi}_{n+1}, D_{n+1,n+1}) = -\infty. 
$$

\end{lemma}

\noindent
Note that the first and third term in the expression for $Q_{Chol,i}$ are non-negative. Hence, 
$Q_{Chol,i} ({\boldsymbol \phi}^i, D_{ii})$ takes the value $-\infty$ if and only if 
$$
({\boldsymbol \phi}^i)^t S_{i-1} {\boldsymbol \phi}_i + 2 ({\boldsymbol \phi}^i)^t S_{\cdot i} + 
S_{ii} = 0 \mbox{ and } D_{ii} = 0. 
$$

\noindent
Let $\mathcal{T}_p$ denote the space of $p \times p$ lower triangular matrices with unit 
diagonal entries, and $\mathcal{D}_p$ denote the space of $p \times p$ diagonal matrices 
with positive diagonal entries. Since $\{({\boldsymbol \phi}^i, D_{ii})\}_{i=1}^p$ forms a 
disjoint partition of $(T,D)$, it follows from Lemma \ref{leminfimum} that if $n < p$, then 
$$
\inf_{T \in \mathcal{T}_p, D \in \mathcal{D}_p} Q_{Chol} (T,D) = \inf_{D_{11} > 0} 
Q_{Chol,1} (D_{11}) + \sum_{i=2}^p \inf_{{\boldsymbol \phi}^i \in \mathbb{R}^{i-1}, D_{ii} 
> 0} Q_{Chol,i} ({\boldsymbol \phi}^i, D_{ii}) = -\infty, 
$$

\noindent
and the infimum can be achieved only if one of the $D_{ii}$'s takes the value zero 
(which is unacceptable as it corresponds to a singular $\Sigma$). Another issue with the 
approach in \cite{HLPL:2006} is that since the function $Q_{Chol,n+1}$ is not a 
jointly convex or even bi-convex in $({\boldsymbol \phi}_{n+1}, D_{n+1,n+1})$, 
existing results in the literature do not provide a theoretical guarantee that the sequence 
of iterates generated by the block coordinatewise minimization algorithm of 
Huang et al. \cite{HLPL:2006} (which alternates between minimizing with respect to 
${\boldsymbol \phi}_{n+1}$ and $D_{n+1,n+1}$) will converge. If the sequence of iterates 
does converge, it is not clear whether the limit is a global minimum or a local minimum. 
Of course, convergence to a local minimum is not desirable as the resulting estimate is 
not in general meaningful, and as described above, convergence to a global minimum 
will imply that the limit lies outside the range of acceptable parameter values. This 
phenomenon is further illustrated in Section \ref{sec:sparsecholconv}. 

Note that the sparsity patterns in $T$ can be associated with a directed acyclic graph 
$G = (V,E)$, where $V = \{1,2, \cdots, p\}$ and $E = \{i \rightarrow j: \; i < j, T_{ij} \neq 
0\}$. Shajoie and Michalidis \cite{Shajoie:Michalidis:2010} use this association to note 
that the problem of choosing a sparsity pattern in $T$ is equivalent to choosing an 
underlying Directed Acyclic Graph (DAG) model. Assuming that $D_{ii} = 1$ for every 
$1 \leq i \leq p$, the authors in \cite{Shajoie:Michalidis:2010} obtain a sparse estimate 
of $T$ by minimizing the objective function 
\begin{equation} \label{eq2}
Q_{Chol} (T, I_p) = tr \left( T^t T S \right) + \lambda \sum_{1 \leq i < j \leq p} |T_{ij}|, 
\end{equation}

\noindent
where $I_p$ denotes the identity matrix of order $p$ (an adaptive lasso version of the 
above objective function is also considered in \cite{Shajoie:Michalidis:2010}). It 
follows from (\ref{eq1}) and (\ref{eq2}) that from an optimization point of view, the 
approach in \cite{Shajoie:Michalidis:2010} is a special case of the approach in 
\cite{HLPL:2006}. Note that fixing $D = I_p$ and only minimizing with respect to 
$T$ significantly simplifies the optimization problem in \cite{HLPL:2006}. Moreover, 
the resulting function in now convex in $T$ with a quadratic term and 
an $\ell_1$ penalty term. The authors in \cite{Shajoie:Michalidis:2010} provide a 
detailed evaluation of the asymptotic properties of their estimator in
an appropriate high-dimensional setting (assuming that $D = I_p$). Owing to 
the interpretation of $\{T_{ij}\}_{j=1}^{i-1}$ as the regression coefficients of $Y_i$ on 
$\{Y_j\}_{j=1}^{i-1}$, this can be regarded as a lasso least squares approach for 
sparsity selection in T. Hence, regardless of whether the true $D_{ii}$'s are all 
equal to one or not, this is a valid approach for model selection/DAG selection, 
which is precisely the goal in \cite{Shajoie:Michalidis:2010}. 

However, we now point out some issues with making the assumption $D_{ii} = 1 \; 
\forall 1 \leq i \leq p$ when the goal is estimation of $\Sigma = T^{-1} D (T^t)^{-1}$. 
Note that if $cov({\bf Y}) = \Sigma$, and if we define the vector of ``latent variables" 
${\bf Z} = T {\bf Y}$, then $cov({\bf Z}) = D$. Hence, assuming that $D_{ii} = 1$ implies 
that the latent variables in ${\bf Z}$ have unit variance, NOT the variables in ${\bf Y}$. 
An assumption of unit variance for ${\bf Y}$ can be dealt with by scaling the 
observations in the data. {\it But scaling the data does not justify the assumption that 
the latent variables in ${\bf Z}$ have unit variances}. This is illustrated in the simulation 
example in Section \ref{sec:simulated}. Also, it is not clear if an assumption of unit 
variances for the latent variables in ${\bf Z}$ can be dealt with by preprocessing the 
data another way. Hence, assuming that the diagonal entries of $D$ are $1$ can be 
restrictive, especially for estimation purposes. 

One could propose an approach where estimates of $T$ are obtained by 
minimizing (\ref{eq2}), and estimates of $D$ are obtained directly from the Cholesky 
decomposition of the sample covariance matrix $S$. However, this approach will not 
work when $n < p$ as $S$ is a singular matrix in this case. To summarize, the 
approach in \cite{Shajoie:Michalidis:2010} is always sensible and useful for the 
purposes of model selection/DAG selection, but makes restrictive assumptions in the 
context of estimation of $(T, D)$. 

In this paper, we develop an $\ell_1$ penalized approach, called Convex Sparse 
Cholesky Selection (CSCS) which provides estimates for $(T, D)$ while inducing 
sparsity in $T$. This approach overcomes the drawbacks of the methods in 
\cite{HLPL:2006} and \cite{Shajoie:Michalidis:2010} while preserving the attractive 
properties of these approaches. The key is to reparameterize in terms of the {\it 
classical} Cholesky parameter for $\Omega$, given by $\Omega = L^tL$. 
In particular, It can be shown that the CSCS objective function is {\it jointly convex} in 
the (nonredundant) entries of $L$, is bounded away from $-\infty$ even if $n < p$, and 
that the sparsity in the classical Cholesky parameter $L$ is exactly reflected in the 
(modified) Cholesky parameter $T$. Furthermore, we provide a cyclic coordinatewise 
minimization algorithm to minimize this objective function, and show that the minimizer 
with respect to each coordinate is unique and can be evaluated in closed form. When 
$n < p$, our objective function in not strictly convex, and convergence of the 
cyclic coordinatewise minimization algorithm does not immediately follow from existing 
results in the literature. We show that recent results in \cite{Khare:Rajaratnam:2014} 
can be adapted in the current context to establish convergence to a global minimum 
for the cyclic coordinatewise minimization algorithm. We show that any global 
minimum lies in the acceptable range of parameter values, i.e., it leads to a positive 
definite estimate of the covariance matrix. We also establish high-dimensional 
asymptotic graph selection and estimation consistency of the resulting 
estimator under standard regularity assumptions. As explained in 
Section \ref{sec:asymptotics}, proving consistency in the current setting is non-trivially 
different than the consistency arguments considered in 
\cite{KOR:2014, PWZZ:2009, Shajoie:Michalidis:2010} because the diagonal entries 
of $L$ are not assumed to be known in this paper. 

A comparison of the relevant properties of the estimators developed in 
\cite{HLPL:2006}, \cite{Shajoie:Michalidis:2010} and this paper is provided in 
Table \ref{table:comparison}. For ease of exposition, we refer to the algorithm in 
\cite{HLPL:2006} as the Sparse Cholesky algorithm, and the one in 
\cite{Shajoie:Michalidis:2010} as the Sparse DAG algorithm. Through experiments 
based on simulated and real datasets, we demonstrate that CSCS can have 
significantly better graph selection as well as estimation performance than Sparse 
Cholesky when $n < p$. These experiments also demonstrate that CSCS can 
improve on the graph selection performance of Sparse DAG, and can lead to 
significant improvements in estimation performance. 

Note that methods inducing sparsity in the Cholesky parameter implicitly 
assume an ordering of the variables. In many applications a natural ordering of the 
variables exists. In the absence of such an ordering, one can employ principled 
methods available in the literature which find the ``best" ordering according to an 
appropriate criterion (see Section \ref{comparison}). 
\begin{table}
 \centering
 \begin{tabular}{|c|c|c|c|}\cline{2-4}
    \multicolumn{1}{c|}{} & \multicolumn{3}{c|}{\bf METHOD} \\ \hline
    {\bf Property}
    & \begin{sideways} Sparse Cholesky \end{sideways}
    & \begin{sideways} Sparse DAG \end{sideways}
    & \begin{sideways} CSCS \end{sideways}
    \\\hline
    {No constraints on sparsity pattern} & + & + & +  \\\hline
    {No constraints on $D$ (for estimation)} & + & & + \\\hline
    {Convergence guarantee to acceptable global minimum when $n < p$} & & + & + 
    \\\hline
    {Asymptotic consistency ($n,p\rightarrow\infty$)} 
    &  & + & + \\\hline
  \end{tabular}
  \caption{Comparison of methods inducing sparsity in the Chloesky parameter of the 
  inverse covariance matrix. Sparse Cholesky refers to the algorithm in \cite{HLPL:2006}, 
  Sparse DAG refers to the algorithm in \cite{Shajoie:Michalidis:2010}. A ``+" indicates 
  that a specified method has the given property. A blank space indicates the absence of 
  a property.}
  \label{table:comparison}
\end{table}

The remainder of the paper is organized as follows. Section \ref{sec:CSCS} introduces 
the CSCS method, and then studies relevant properties such as convergence, 
computational complexity. In Section \ref{comparison}, we compare and contrast the 
CSCS method (which induces sparsity in $T$) with penalized methods which induce 
sparsity in $\Omega$. Section \ref{sec:experiments} illustrates the performance of the 
CSCS method on simulated and real data. Section \ref{sec:asymptotics} establishes 
high-dimensional asymptotic consistency (both estimation and model selection) of the 
CSCS method. The supplementary document contains proofs of some of the results in 
the paper.

\section{A convex approach for sparse Cholesky estimation} \label{sec:CSCS}

\noindent
As pointed out in Lemma \ref{leminfimum}, if $n < p$, the infimum of $Q_{Chol,n+1} 
({\boldsymbol \phi}_{n+1}, D_{n+1,n+1})$ over the range of acceptable values 
of $({\boldsymbol \phi}_{n+1},D_{n+1,n+1})$ is $-\infty$. However, the infimum is 
attained only if $D_{n+1,n+1} = 0$, which is outside the range of acceptable values of 
$D_{n+1,n+1}$. Also, since $Q_{Chol} (T,D)$ is not jointly convex in $(L,D)$, their are 
no convergence guarantees for the block coordinatewise minimization algorithm proposed 
in \cite{HLPL:2006}. Given the attractive properties of convex functions and the rich theory 
for convex optimization, a natural approach to address these issues is to develop a convex 
objective function for this problem. Such an approach will also potentially lead to a deeper 
theoretical analysis of the properties of the solution and corresponding algorithm. 
The objective function $Q_{Chol} (T, I_p)$ used in \cite{Shajoie:Michalidis:2010} is jointly 
convex in $T$, but we want to avoid any restrictive constraints on $D$. 

\subsection{The CSCS objective function}

\noindent
We will now show that all the goals mentioned above can be achieved by reparametrizing in 
terms of the classical Cholesky parameter. Recall that the classical Cholesky decomposition 
of $\Omega$ is given by $\Omega = L^t L$, where $L$ (which we will refer to as the 
classical Cholesky parameter) is a lower triangular matrix with positive diagonal entries. It is 
easy to see that 
\begin{equation} \label{eq3}
L_{ij} = T_{ij}/\sqrt{D_{jj}} \mbox{ for every } i \leq j. 
\end{equation}

\noindent
Hence, $L_{ij} = 0$ if and only if $T_{ij} = 0$, i.e., {\it sparsity in $T$ is equivalent to 
sparsity in $L$}. After reparametrizing $Q_{Chol}$ in terms of $L$ (as opposed to 
$(T,D)$) and some simple manipulations, we obtain the following objective function. 
\begin{equation} \label{eq4}
Q_{Chol} (T) = tr \left( LL^t S \right) - 2 \log |L| + \lambda \sum_{1 \leq j < i \leq p} 
|L_{ij}| L_{jj}. 
\end{equation}

\noindent
Note that the first term in (\ref{eq4}) is a quadratic form in the entries of $L$, and hence is 
jointly convex in the entries of $L$. Since $L$ is a lower triangular matrix, it follows that 
$- \log |L| =  \sum_{i=1}^p -\log L_{ii}$, and hence the second term in (\ref{eq4}) is also 
jointly convex in entries of $L$. However, terms of the form $|L_{ij}| L_{jj}$ are not 
jointly convex, and hence the penalty term in (\ref{eq4}) is not jointly convex either. 
Hence, we replace the penalty term $\lambda \sum_{1 \leq j < i \leq p} |L_{ij}| L_{jj}$ by 
the term $\lambda \sum_{1 \leq j < i \leq p} |L_{ij}|$ (which is jointly convex in the entries 
of $L$), and introduce the following objective function. 
\begin{equation} \label{eq5}
Q_{CSCS} (L) = tr \left( L^t L S \right) - 2 \log |L| + \lambda \sum_{1 \leq j < i \leq p} 
|L_{ij}|. 
\end{equation}

\noindent
The following lemma immediately follows from (\ref{eq3}) and the discussion above. 
\begin{lemma}[Joint convexity] \label{lem:convexity}
$Q_{CSCS} (L)$ is jointly convex in the entries of $L$. Also, a global minimizer of $Q_{CSCS}$ will 
be sparse in $L$ (and hence sparse in $T$). 
\end{lemma}

\noindent
Let ${\boldsymbol \eta}^i = (L_{ij})_{j=1}^i$ denote the vector of lower triangular and 
diagonal entries in the $i^{th}$ row of $L$ for $1 \leq i \leq p$. Recall that $S_i$ denotes 
the $i \times i$ sub matrix of $S$ starting from the first row (column) to the $i^{th}$ row 
(column). Let $L_{i \cdot}$ denote the $i^{th}$ row of $L$, for $1 \leq i \leq p$. 
It follows from (\ref{eq5}), the lower triangular nature of $L$, and the definition of 
${\boldsymbol \eta}^i$ that 
\begin{eqnarray}
Q_{CSCS} (L) 
&=& tr \left( L S L^t \right) - 2 \sum_{i=1}^p \log L_{ii} + \lambda \sum_{1 \leq j < i \leq p} 
|L_{ij}| \nonumber\\
&=& \sum_{i=1}^p L_{i \cdot} S L_{i \cdot}^t - 2 \sum_{i=1}^p \log 
\eta^i_i + \lambda \sum_{i=2}^p \sum_{j=1}^{i-1} 
|\eta^i_j| \nonumber\\
&=& \sum_{i=1}^p ({\boldsymbol \eta}^i)^T S_i {\boldsymbol \eta}^i - 2 \sum_{i=1}^p 
\log \eta^i_i + \lambda \sum_{i=2}^p \sum_{j=1}^{i-1} 
|\eta^i_j| \nonumber\\
&=& \sum_{i=1}^p Q_{CSCS,i} ({\boldsymbol \eta}^i), \label{eq6}
\end{eqnarray}

\noindent
where 
\begin{equation} \label{eq7}
Q_{CSCS,i} ({\boldsymbol \eta}^i) = ({\boldsymbol \eta}^i)^T S_i {\boldsymbol \eta}^i - 2 
\log \eta^i_i + \lambda \sum_{j=1}^{i-1} |\eta^i_j| 
\end{equation}

\noindent
for $2 \leq i \leq p$, and 
\begin{equation} \label{eq8}
Q_{CSCS,1} (L_{11}) = L_{11}^2 S_{11} - 2 \log L_{11}. 
\end{equation}

\noindent
Let $\mathcal{L}_p$ denote the space of $p \times p$ lower triangular matrices with positive 
diagonal entries. Our next goal is to establish that the function $Q_{CSCS} (L)$ is uniformly 
bounded below over $L$. We will assume that the diagonal entries of the sample covariance 
matrix $S$ are strictly positive. This basically means that none of the underlying $p$ 
marginal distributions are degenerate. We now state a lemma from 
\cite{Khare:Rajaratnam:2014} which will play a crucial role in this exercise. 
\begin{lemma}[\cite{Khare:Rajaratnam:2014}] \label{lowerbound}
Let $A$ be a $k \times k$ positive semi-definite matrix with $A_{kk} > 0$, and $\lambda$ be a 
positive constant. Consider the function 
$$
h ({\bf x}) = - \log x_k + {\bf x}^T A {\bf x} + \lambda \sum_{i=1}^{k-1} |x_j| 
$$

\noindent
defined on $\mathbb{R}^{k-1} \times \mathbb{R}_+$. Then, there exist positive constants $a_1$
and $a_2$ (depending only on $\lambda$ and $A$), such that 
$$
h ({\bf x}) \geq a_1 x_k - a_2 
$$

\noindent
for every ${\bf x} \in \mathbb{R}^{k-1} \times \mathbb{R}_+$. 
\end{lemma}

\noindent
Using (\ref{eq7}), (\ref{eq8}) along with the facts that $S_i$ is positive semi-definite and $S_{ii} > 0$, 
it follows from Lemma \ref{lowerbound} that for every $1 \leq i \leq p$, there exist positive constants 
$a_i$ and $b_i$ such that 
\begin{eqnarray}
Q_{CSCS,i} ({\boldsymbol \eta}^i) 
&=& ({\boldsymbol \eta}^i)^T S_i {\boldsymbol \eta}^i - 2 \log \eta^i_i + \frac{\lambda}{2} \sum_{j=1}^{i-1} 
|\eta^i_j| + \frac{\lambda}{2} \sum_{j=1}^{i-1} |\eta^i_j| \nonumber\\
&\geq& a_i \eta^i_i - b_i + \frac{\lambda}{2} \sum_{j=1}^{i-1} |\eta^i_j| \label{eq9}
\end{eqnarray}

\noindent
for every ${\boldsymbol \eta}^i \in \mathbb{R}^{i-1} \times \mathbb{R}_+$. The following lemma now follows 
immediately from (\ref{eq6}), (\ref{eq9}) and the fact that $\{{\boldsymbol \eta}^i\}_{i=1}^p$ forms a disjoint 
partition of $L$. 
\begin{lemma} \label{lemlwrbd}
For every $n$ and $p$, 
$$
\inf_{L \in \mathcal{L}_p} Q_{CSCS} (L) = \sum_{i=1}^p \inf_{{\boldsymbol \eta}^i \in 
\mathcal{R}^{i-1} \times \mathcal{R}_+} Q_{CSCS,i} ({\boldsymbol \eta}^i) \geq -
\sum_{i=1}^p b_i > -\infty, 
$$

\noindent
and $Q_{CSCS} (L) \rightarrow \infty$ as $|\eta^i_j| = |L_{ij}| \rightarrow \infty$ for any 
$j < i$, or as $\eta^i_i = L_{ii} \rightarrow 0$. Hence, any global minimum of 
$Q_{CSCS,i}$ has a strictly positive value for $\eta^i_i = L_{ii}$, and hence {\bf any 
global minimum of $Q_{CSCS}$ over the open set $\mathcal{L}_p$ lies in 
$\mathcal{L}_p$}. 
\end{lemma}

\subsection{A minimization algorithm for $Q_{CSCS}$}

\noindent
We now provide an algorithm to minimize the convex objective function $Q_{CSCS} (L)$. Since 
$\{{\boldsymbol \eta}^i\}_{i=1}^p$ form a disjoint partition of the (nonredundant) parameters in $L$, 
it follows that optimizing $Q_{CSCS} (L)$ is equivalent to separately optimizing $Q_{CSCS,i} 
({\boldsymbol \eta}^i)$ for $1 \leq i \leq p$. 

Consider, similar to Lemma \ref{lowerbound}, a generic function of the form 
\begin{equation} \label{eq10}
h_{k,A,\lambda} ({\bf x}) = - 2 \log x_k + {\bf x}^T A {\bf x} + \lambda \sum_{i=1}^{k-1} |x_j| 
\end{equation}

\noindent
from $\mathbb{R}^{k-1} \times \mathbb{R}_+$ to $\mathbb{R}$. Here $k$ is a positive integer, 
$\lambda > 0$, and $A$ is a positive semi-definite matrix with positive diagonal entries. It 
follows from (\ref{eq7}) and (\ref{eq8}) that $Q_{CSCS,i} ({\boldsymbol \eta}^i) = h_{i,S_i,\lambda} 
({\boldsymbol \eta}^i)$ for every $1 \leq i \leq p$. It therefore suffices to develop an algorithm to 
minimize a function of the form $h_{k,A,\lambda}$ as specified in (\ref{eq10}). Note that without the 
logarithmic term and the restriction that $x_k > 0$, the optimization problem for $h_{k,A,\lambda}$ 
would have been equivalent to the lasso optimization problem for which several approaches 
have been developed in the literature, such as the shooting algorithm in \cite{Fu:1998}, or the 
pathwise coordinate optimization approach in \cite{FHT:2008b}, for example. However, these 
algorithms do not apply in the current situation due to the presence of the logarithmic term and 
the condition $x_k > 0$. 

We will now derive a cyclic coordinatewise minimization algorithm for $h_{k,A,\lambda}$. For 
every $1 \leq j \leq k$, define the function $T_j: \mathbb{R}^{k-1} \times \mathbb{R}_+ 
\rightarrow \mathbb{R}^{k-1} \times \mathbb{R}_+$ by 
\begin{equation} \label{eq11}
T_j ({\bf x}) = \inf_{{\bf y} \in \mathbb{R}^{k-1} \times \mathbb{R}_+: y_l = x_l \forall l \neq j} 
h_{k,A,\lambda} ({\bf x}). 
\end{equation}

\noindent
The following lemma (proof provided in the supplemental document) shows that the 
functions $\{T_j\}_{j=1}^k$ can be computed in closed form. 
\begin{lemma} \label{closedform}
The function $T_j ({\bf x})$ defined in (\ref{eq11}) can be computed in closed form. In particular, 
\begin{equation} \label{eq12}
\left( T_j ({\bf x}) \right)_j = \frac{S_{\lambda} \left(- 2 \sum_{l \neq j} A_{lj} x_l \right)}{2 A_{jj}}
\end{equation}

\noindent
for $1 \leq j \leq k-1$, and 
\begin{equation} \label{eq13}
(T_k ({\bf x}))_k = \frac{- \sum_{l \neq k} A_{lk} x_l  + \sqrt{\left( \sum_{l \neq k} A_{lk} x_l 
\right)^2 + 4 A_{kk}}}{2 A_{kk}}. 
\end{equation}
\end{lemma}

\noindent
Here $S_\lambda$ is the soft-thresholding operator given by $S_\lambda (x) = sign(x)(|x| - 
\lambda)_+$. Lemma \ref{closedform} provides the required ingredients to construct a cyclic 
coordinatewise minimization algorithm to minimize $h_{k,A,\lambda}$ (see 
Algorithm \ref{algorithm1}). Now, to minimize $Q_{CSCS} (L)$, we use Algorithm \ref{algorithm1} 
to minimize $Q_{CSCS,i} ({\boldsymbol \eta}^i)$ for every $1 \leq i \leq p$, and combine the 
outputs to obtain the a matrix on $\mathcal{L}_p$ (see Algorithm \ref{algorithm2}). We refer to 
Algorithm \ref{algorithm2} as the CSCS algorithm. 
\begin{figure}
  \centering
  \begin{minipage}[t]{0.8\textwidth}
    \alglanguage{pseudocode}
    \begin{algorithm}[H]
      \caption{(Cyclic coordinatewise algorithm for $h_{k,A,\lambda}$)} \label{algorithm1}
      \begin{algorithmic}
        \State Input: $k$, $A$ and $\lambda$
        \State Input: Fix maximum number of iterations: $r_{max}$
        \State Input: Fix initial estimate: $\hat{\bf x}^{(0)}$
        \State Input: Fix convergence threshold: $\epsilon$
        \State Set $r \gets 1$
        \State converged = FALSE
        \State Set $\hat{\bf x}^{\cur} \gets \hat{\bf x}^{(0)}$
        \Repeat
        
        \State $\hat{\bf x}^{\old} \gets \hat{\bf x}^{\cur}$
        \Statex 
        \For{$j \gets 1, 2, \cdots, k-1$}
        \begin{align*}
          \hat{x}^{\cur}_j \gets (T_j ({\bf x}^{\cur}))_j
          \end{align*}
         \EndFor
        
        \begin{align}
          \hat{x}^{\cur}_k \gets (T_k ({\bf x}^{\cur}))_k
        \end{align}

        \Statex
        
        \Statex
        \State $\hat{\bf x}^{(r)} \gets \hat{\bf x}^{\cur}$
        \Statex\hspace{\algorithmicindent}{\tt \#\# Convergence checking}
        \If{$\|\hat{\bf x}^{\cur} - \hat{\bf x}^{\old}\|_\infty < \epsilon$}
        \State converged = TRUE
        \Else
        \State $r \gets r + 1$
        \EndIf
        \Statex

        \Until{converged = TRUE or $r > r_{\max}$}
        \State Return final estimate: $\hat{\bf x}^{(r)}$ 
      \end{algorithmic}
    \end{algorithm}
  \end{minipage}  
\end{figure}

\begin{figure}
  \centering
  \begin{minipage}[t]{0.8\textwidth}
    \alglanguage{pseudocode}
    \begin{algorithm}[H]
      \caption{(CSCS algorithm: minimization algorithm for $Q_{CSCS}$)} \label{algorithm2}
      \begin{algorithmic}
        \State Input: Data ${\bf Y}_1, {\bf Y}_2, \cdots, {\bf Y}_n$ and $\lambda$
        \State Input: Fix maximum number of iterations: $r_{max}$
        \State Input: Fix initial estimate: $\hat{L}^{(0)}$
        \State Input: Fix convergence threshold: $\epsilon$
        
        \Statex 
        \For{$i \gets 1, 2, \cdots, p$}
        \State $(\hat{{\boldsymbol \eta}^i})^{(0)}$ $\gets$ $i^{th}$ row of $\hat{L}^{(0)}$ (up to the 
        diagonal) 
        \State Set $\hat{{\boldsymbol \eta}^i}$ to be minimizer of $Q_{CSCS,i}$ obtained by using 
        Algorithm \ref{algorithm1} 
        \State \hspace{0.16in} with $k=i, A = S_i, \lambda, r_{max}, \hat{\bf x}^{(0)} = 
        (\hat{{\boldsymbol \eta}^i})^{(0)}, \epsilon$ 
        \EndFor
        
        \Statex
        
        \State Construct $\hat{L} \in \mathcal{L}_p$ by setting its $i^{th}$ row (up to the diagonal) 
        as $\hat{{\boldsymbol \eta}^i}$ 
        \State Return final estimate: $\hat{L}$ 
      \end{algorithmic}
    \end{algorithm}
  \end{minipage}  
\end{figure}

Note that although the function $Q_{CSCS,i}$ is jointly convex in the entries of 
${\boldsymbol \eta}^i$, it is not in general strictly convex if  $n < i$, and does not necessarily 
have a unique global minimum. Hence, it is not immediately clear if existing results in the 
literature imply the convergence of Algorithm \ref{algorithm2} to a global minimum of 
$Q_{CSCS}$. The next theorem invokes results in \cite{Khare:Rajaratnam:2014} to establish 
convergence of Algorithm \ref{algorithm2}. 
\begin{thm} \label{thm:convergence}
If $S_{ii} > 0$ for every $1 \leq i \leq p$, then Algorithm \ref{algorithm2} converges to a global 
minimum of $Q_{CSCS}$. 
\end{thm}

\noindent
The proof of the above theorem is provided in the supplemental document. 

\subsection{Selection of tuning parameter} \label{sec:tuning}

\noindent
The tuning parameter $\lambda$ can be selected using a "BIC"-like measure, 
defined as follows: 
$$
BIC(\lambda) = n \text{tr}(S\hat{\Omega}) - n \log |\hat{\Omega}| + \log n*E
$$
where $E$ denotes the number of non-zero entries in $\hat{L}$, $n$ is
the sample size, $S$ the sample covariance and 
$\hat{\Omega} = \hat{L}^t \hat{L}$. The value of $\lambda$ minimizing the 
function $BIC(\lambda)$ can be chosen. 

In \cite{HLPL:2006} and \cite{Shajoie:Michalidis:2010} the authors respectively 
propose tuning parameter choices based on cross-validation and scaled normal
quantiles. These procedures are described briefly in Section \ref{sec:applicationcall} 
and Section \ref{sec:simulated} respectively. 

\subsection{Computational complexity of the CSCS algorithm}

\noindent
We now proceed to evaluate the computational complexity of the CSCS algorithm. Note that 
the CSCS algorithm (Algorithm \ref{algorithm2}) involves $p$ separate minimizations, all 
of which can be run in parallel, especially given modern computing resources. In a 
parallelizable setting, we define the computational complexity as a maximum number of 
computations among all processes running in parallel. We will show the following. 
\begin{lemma} \label{complexity}
The best case computational complexity per iteration for Algorithm \ref{algorithm2} is\\ 
$\min(O(np), O(p^2))$ (if all the $p$ minimizations are run in parallel), and the worst case 
computational complexity per iteration for Algorithm \ref{algorithm2} is $\min \left( O \left( n 
\sum_{i=1}^p i \right), O \left( \sum_{i=1}^n i^2 \right) \right) = \min(O(np^2, p^3))$ (if all the 
$p$ minimizations are run sequentially). 
\end{lemma}

\noindent
To prove the above lemma, we start by establishing a result about the computational 
complexity per iteration for Algorithm \ref{algorithm1}. 
\begin{lemma} \label{cmxbasic}
Suppose $A = BB^T$, where $B$ is an $k \times n$ matrix. Then the computational 
complexity for Algorithm \ref{algorithm1} is $\min(O(nk), O(k^2))$. 
\end{lemma}

\noindent
The proof of this lemma is provided in the appendix. Since $S = \frac{1}{n} \sum_{j=1}^n 
{\bf Y}_j {\bf Y}_j^T$, it follows that $S_i$ (a principal $i \times i$ submatrix of $S$) can 
be written as $B_i B_i^T$ for an appropriate $i \times n$ matrix $B_i$. Since 
$Q_{CSCS,i} ({\boldsymbol \eta}^i) = h_{i,S_i,\lambda} ({\boldsymbol \eta}^i)$ for every 
$1 \leq i \leq p$, Lemma \ref{complexity} follows immediately by invoking 
Lemma \ref{cmxbasic}. 

\subsection{Comparison and connections with penalized sparse partial correlation 
methods} \label{comparison}

\noindent
In this section we compare and contrast the CSCS method (which induces sparsity in 
the Cholesky factor of $\Omega$) with sparse partial correlation methods, i.e., 
penalized methods which induce sparsity in the inverse covariance matrix $\Omega$ 
itself. The entries in the $i^{th}$ row of $\Omega$ (appropriately scaled) can be 
interpreted as regression coefficients of the $i^{th}$ variable against {\it all} other 
variables. Recall that the (non-redundant) entries in the $i^{th}$ row of $T$, on the 
other hand, are the regression coefficients of the $i^{th}$ variable against only the {\it 
preceding} variables. A natural question to ask is whether there is any connection 
between models which introduce sparsity in the Cholesky factor of $\Omega$ and 
models which induce sparsity in $\Omega$ itself.  In general, the sparsity pattern in 
the Cholesky factor $T$ of a positive definite matrix $\Omega$ is not the same as the 
sparsity pattern in $\Omega$ itself. Note that a given pattern of zeros in the lower 
triangle a $p \times p$ matrix uniquely corresponds to a graph with vertices $\{1,2, 
\cdots, p\}$, where two vertices do not share an edge whenever the corresponding 
entry is included in the pattern of zeros. It is known that the sparsity pattern in 
$\Omega$ is exactly the same as its Cholesky factor if and only if the corresponding 
graph is chordal (decomposable) and the vertices are ordered based on a perfect 
vertex elimination scheme (see \cite{PSS:1989}). 

We now summarize the relevant details of penalized methods which induce sparsity in 
$\Omega$. Such methods can be divided into two categories: penalized likelihood methods 
such as GLASSO (\cite{BED:2008}, \cite{FHT:2008a}), and penalized pseudo-likelihood 
methods such as CONCORD (\cite{KOR:2014}), SPACE (\cite{PWZZ:2009}) and 
SYMLASSO (\cite{FHT:2010}).  The GLASSO objective function is comprised of a log 
Gaussian likelihood term and an $\ell_1$-penalty term for entries of $\Omega$. 
Friedman et al. \cite{FHT:2008a} present an algorithm for minimizing this objective 
function with has computational complexity of $O(p^3)$ per iteration~\footnote{In recent 
years, several adaptations/alternatives to this algorithm have been proposed in order to 
improve its speed (see \cite{HSDR:2011, Mazumder:Hastie:2012} for instance). However, 
for these methods to provide substantial improvements over the graphical lasso, certain 
assumptions are required on the number and size of the connected components of the 
graph implied by the zeros in the minimizer.}. Pseudo-likelihood based objective functions 
used in CONCORD, SPACE and SYMLASSO are comprised of a log pseudo-likelihood 
trem which is based on the regression based interpretation of the entries of $\Omega$, 
and an $\ell_1$-penalty term for entries of $\Omega$. These objective functions are 
typically minimized using cyclic coordinatewise minimization with a computational 
complexity of $\min(O(np^2), O(p^3))$~\footnote{Recently, a much faster proximal 
gradient based optimization method for the CONCORD objective function has been 
developed in \cite{ODKR:2014}.}. Owing to the regression based interpretation of the 
pseudo-likelihood, the minimization is done over all symmetric matrices with positive 
diagonal entries (as opposed to GLASSO, where the minimization is done over the set of 
positive definite matrices), and hence the minimizer is not guaranteed to be positive definite. 
In many applications, the main goal is selection of the sparsity pattern (network), and this 
does not pose a problem. In fact, getting rid of the positive definiteness constraint is 
helpful in improving the performance of such methods (as compared to GLASSO) in 
high-dimensional settings (see \cite{KOR:2014}). The CONCORD algorithm, unlike 
SPACE and SYMLASSO, provides crucial theoretical guarantees of convergence to a 
global minimum of the respective objective function (while preserving all the other attractive 
properties of SPACE and SYMLASSO). 

There is, in fact, an interesting parallel between CONCORD and CSCS. The CONCORD objective 
function (scaled by $\frac{2}{n}$) is given by 
$$
Q_{\con} (\Omega) = -\sum_{i=1}^p 2 \log \omega_{ii} + tr \left( \Omega^t \Omega S \right) + 
\lambda \sum_{1 \leq j < i \leq p} |\omega_{ij}|. 
$$

\noindent
On the other hand, it follows from (\ref{eq5}) that the CSCS objective function can be written as 
$$
Q_{CSCS} (L) = -\sum_{i=1}^p 2 \log L_{ii} + tr \left( L^t L S \right) + \lambda \sum_{1 \leq j < i 
\leq p} |L_{ij}|. 
$$

\noindent
{\it Hence, from a purely mathematical point of view, CONCORD and CSCS are both maximizing the 
same objective function. The difference is that CONCORD optimizes the function over the set of 
symmetric matrices with positive diagonal entries, whereas CSCS optimizes the function over the 
set of lower triangular matrices with positive diagonal entries.} Despite this very close connection 
between the objective functions for CONCORD and CSCS, the difference in the range of 
optimization leads to some qualitative differences between the respective 
optimization algorithms and estimators. 
\begin{enumerate}[(a)]
\item (Computational Complexity) The parallelizability of the $p$ minimizations in the 
CSCS algorithm, gives it a distinct computational advantage over the CONCORD 
algorithm (which is not parallelizable). Even in the worst case, when all the $p$ 
minimizations for CSCS are implemented sequentially, the computational complexity is 
the same as CONCORD (by Lemma \ref{complexity}). 
\item (Positive definiteness of resulting estimator of $\Omega$) As discussed above, the 
CONCORD estimator (and other pseudo-likelihood based estimators) for $\Omega$ is not 
guaranteed to be positive definite. However, the estimator for $\Omega$ constructed by 
taking the CSCS estimator and multiplying it by its transpose, is always positive definite. 
\item (Ordering of variables) The CSCS algorithm uses an implicit ordering of the variables, 
whereas the CONCORD algorithm (as well as GLASSO, SPACE and SYMLASSO) do 
not need such an ordering. While this does not pose a problem for CSCS in applications where 
there is a natural ordering of variables, a principled method is needed for the choice of ordering in 
other applications. Two such methods have been recently developed in 
\cite{Dellaportas:Pourahmadi:2012, Rajaratnam:Salzman:2013}. 
\end{enumerate}


\noindent
We close this section by observing that as discussed above, the regression based 
interpretation for the entries of $\Omega$ leads to a different objective function than 
the log Gaussian likelihood for $\Omega$. However, it can be easily shown that the 
objective function based on the regression based interpretation for the entries of the 
Cholesky factor $T$ (or equivalently $L$) exactly corresponds to the log Gaussian 
likelihood for $T$.

\section{Experiments} \label{sec:experiments}

\subsection{Sparse Cholesky convergence when $n < p$} \label{sec:sparsecholconv}

\noindent
In this section we illustrate that when $n < p$, the Sparse Cholesky algorithm in 
\cite{HLPL:2006} can converge to a limit where at least one of the $D_{ii}$'s takes 
the value zero. As discussed in the introduction, such a limit corresponds to a 
singular $\Sigma$ and lies outside the range of acceptable parameter
values. It is quite common to find situations where this happens, and
we provide such an example below. 

\begin{figure}[H]
\begin{center}
  \includegraphics[scale=0.4]{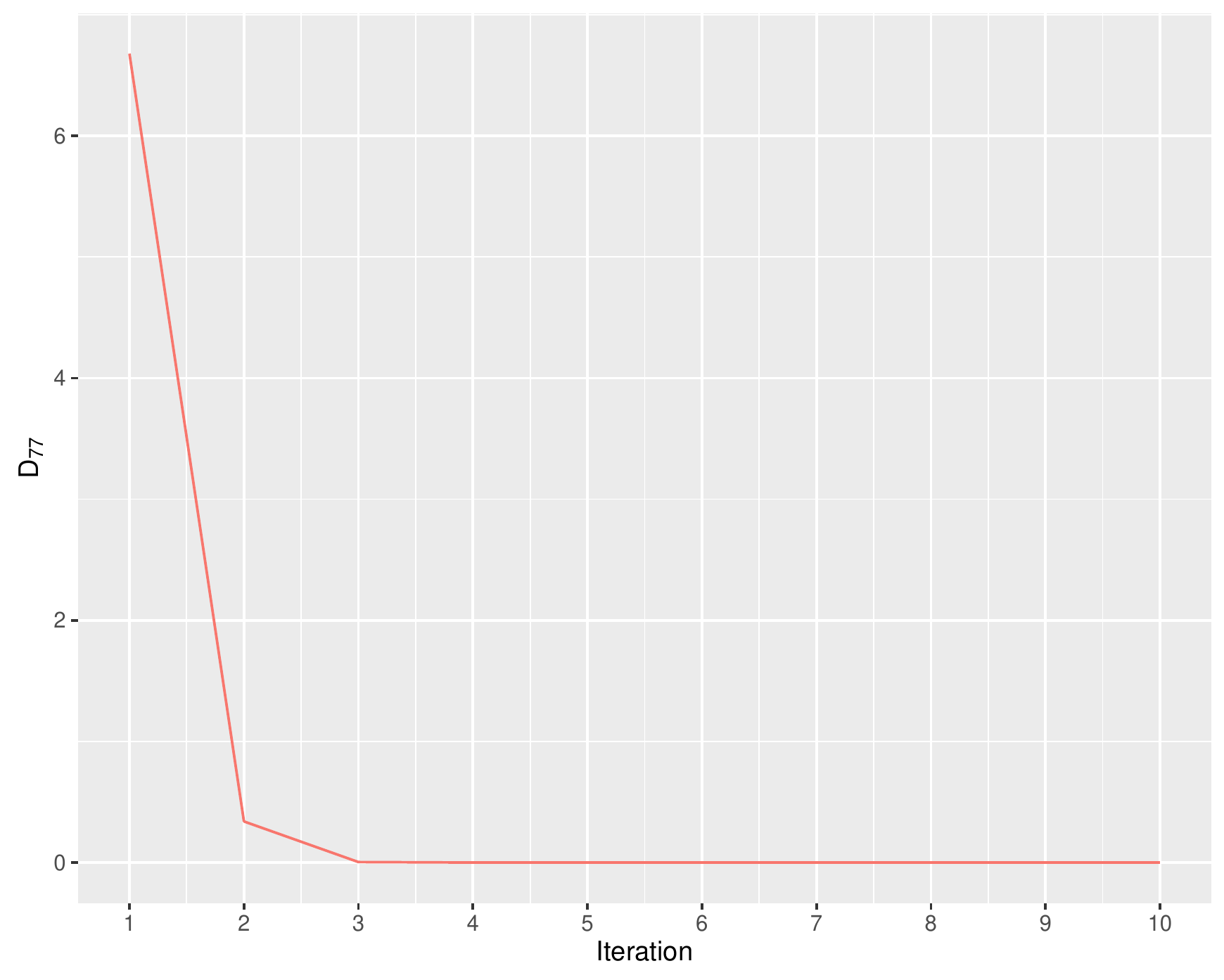}
\end{center}
\caption{Plot of the iterates for $D_{77}$ for Sparse Cholesky in a setting with $p = 8$. 
It shows how the value jumps to 0 (and stays there).} 
\label{fig:d77}
\end{figure}

We chose $p=8$ and generated $\Omega_0 = T_0^t D_0^{-1} T_0$ in the
following manner. Sixty percent of the lower 
triangular entries of $T_0$ are randomly set to zero. The remaining
$40 \%$ entries are chosen 
from a uniform distribution on $[0.3,0.7]$ and then assigned a 
positive/negative sign with probability $0.5$. Now, a $p \times p$ diagonal matrix 
$D_0$ is generated with diagonal entries chosen uniformly from
$[2,5]$. We then set $n = p-1$ and generate data from the multivariate normal 
distribution with mean ${\bf 0}$ and covariance matrix $\Omega_0$. We 
initialize $T$ and $D$ to be $I_8$, and run the Sparse Cholesky algorithm. 
After $4$ interations, $D_{77}$ jumps to 0 and stays there, as shown in 
Figure \ref{fig:d77}. This leads to a degenerate covariance matrix estimate. 

\subsection{Simulated data: Graph Selection and Estimation} \label{sec:simulated}

\noindent
In this section, we perform a simulation study to compare the graph/model 
selection and estimation performance of CSCS, Sparse Cholesky and Sparse DAG. 
For model selection, we consider eight different settings with $p = 1000, 2000$ and 
$n = p/8, p/4, p/2, 3p/2$. In particular, for each $p \in \{1000, 2000\}$, a $p \times p$ 
lower triangular matrix $T_0$ is generated as follows. We randomly choose 
$98 \%$ of the lower triangular entries, and set them to zero. The remaining 
$2 \%$ entries are chosen randomly from a uniform distribution on $[0.3,0.7]$ and 
then assigned a positive/negative sign with probability $0.5$. Now, a $p \times p$ 
diagonal matrix $D_0$ is generated with diagonal entries chosen uniformly from 
$[2,5]$. For each sample size $n = p/8, p/4, p/2, 3p/2$, $100$ datasets, each 
having i.i.d. multivariate normal distribution with mean zero and inverse covariance 
matrix $\Omega_0 = T_0^t D_0^{-1} T_0$, are generated. 

The model selection performance of the three algorithms, CSCS, Sparse 
Cholesky, Sparse DAG, is then compared using receiver operating characteristic 
(ROC) curves. These curves compare true positive rates (TPR) and false positive 
rates (FPR), and are obtained by varying the penalty parameter over roughly $40$ 
possible values. In applications, FPR is typically controlled to be sufficiently small, 
and therefore we restrict ourselves to settings where the FPR is less than 0.15. 
Area-under-the-curve is a standard measure used to compare model selection 
performance (see \cite{Fawcett2006}, \cite{FHT:2010}). 

Tables \ref{tbl:aucp1000} and \ref{tbl:aucp2000} show the mean and standard 
deviation (over 100 simulations) for the area-under-the-curve for CSCS, Sparse 
Cholesky and Sparse DAG for $p = 1000,2000$ and $n = p/8, p/4, p/2, 3p/2$. It is 
clear that CSCS has a better model selection performance as compared to Sparse 
Cholesky and Sparse DAG for all the settings. 
\begin{enumerate}[(a)]
\item As expected Sparse Cholesky performs significantly worse that the other 
methods when $n<p$, but its comparative (and absolute) performance improves with 
increasing sample size, especially when $n > p$. 
\item The tables also show that CSCS does better than Sparse DAG in terms of model 
selection, although the difference in AUC is not as drastic as with Sparse 
Cholesky. In should be noted that {\it CSCS has a higher AUC than Sparse DAG for 
each of the $800$ datasets (100 each for $p = 1000,2000$ and $n = p/8, p/4, p/2, 
3p/2$)}. We also note that the variability is much lower for CSCS than the other 
methods. 
\end{enumerate}

It is worth mentioning that for each of the $800$ datasets, the data was centered and 
scaled before running each method. This is done firstly to illustrate that
scaling the data does not justify assuming that the latent variable
conditional variances $\{D_{ii}\}_{i=1}^p$ are identically $1$, borne
out by the consistently better model selection performance of CSCS as
opposed to Sparse DAG. Secondly, we observed that the three algorithms 
typically run much faster when the data is scaled. Also, premultiplication of 
a multivariate normal vector by a diagonal matrix does not affect the sparsity pattern 
in the Cholesky factor of the inverse. Hence, given the extensive nature of our 
simulation study, we scaled the data in the interest of time. 

\begin{figure}[H]
  \begin{subfigure}[b]{0.45\textwidth}
    \includegraphics[width=\textwidth]{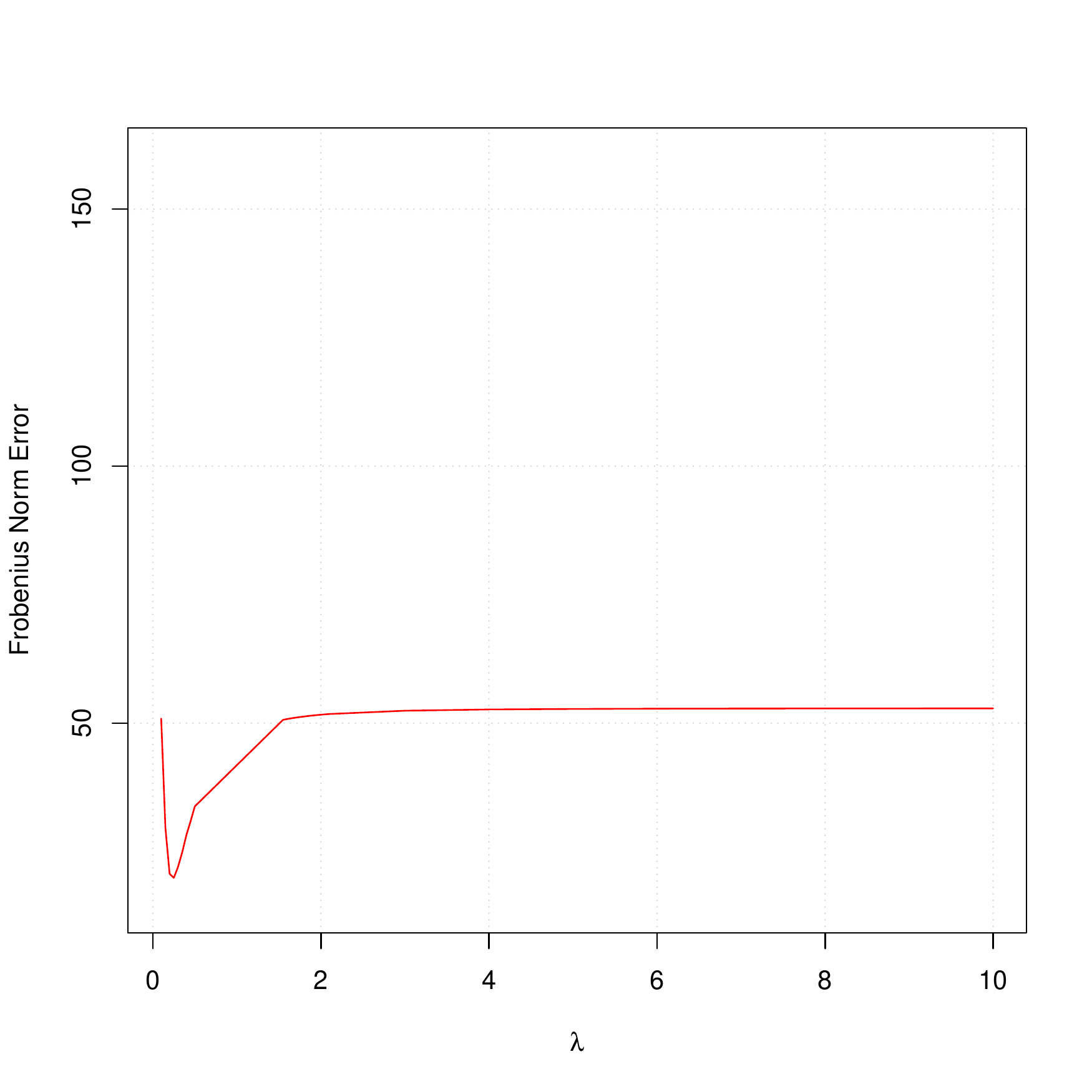}
    \caption{Frobenius Norm Error for CSCS\\
    (y-axis) with varying penalty parameter\\ 
    value (x-axis) for $n = 500$}
    \label{fig:cscsfrob500}
  \end{subfigure}
  \begin{subfigure}[b]{0.45\textwidth}
    \includegraphics[width=\textwidth]{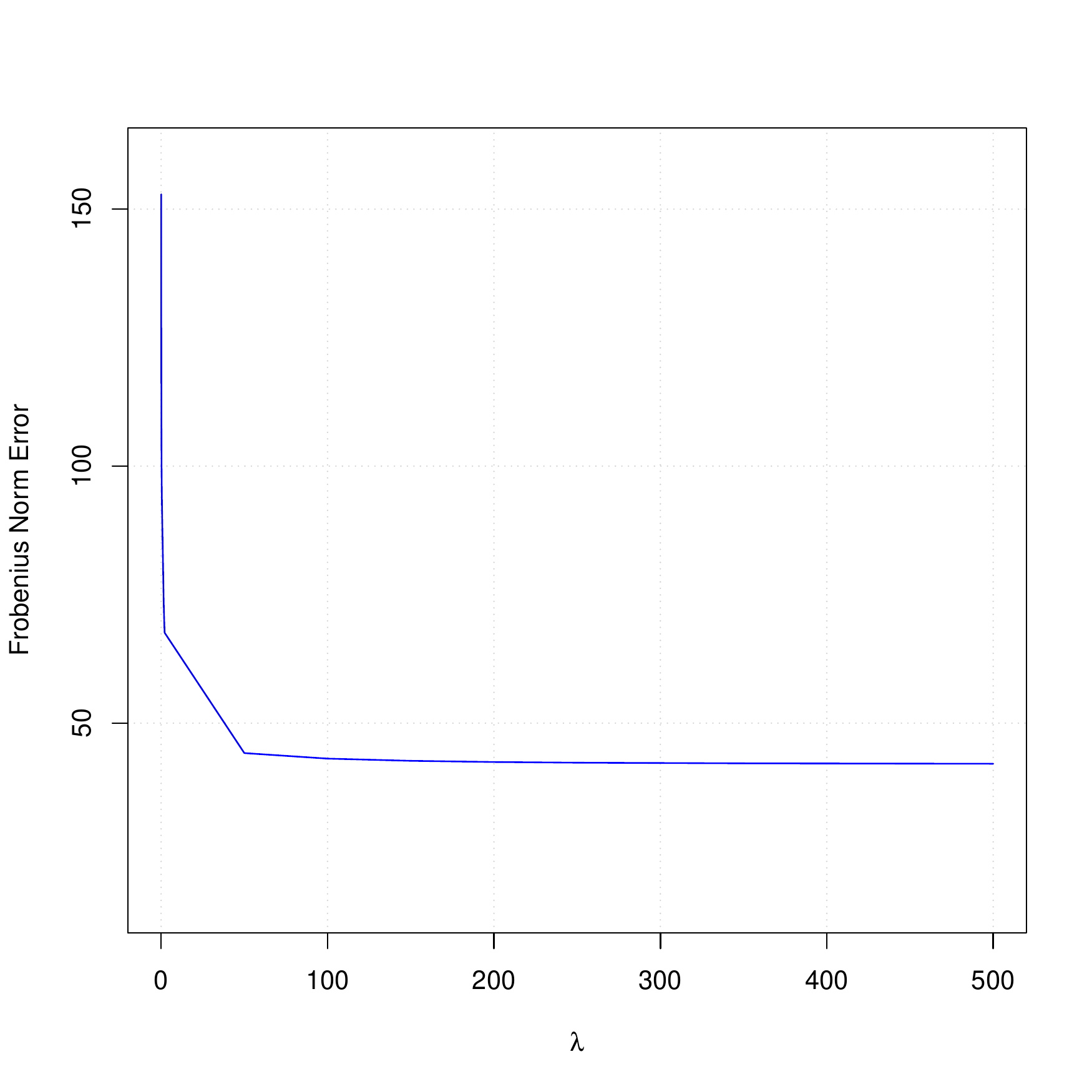}
    \caption{Frobenius Norm Error for 
Sparse DAG averaged over 50 replications for $n = 500$ for different penalty 
parameter values.}
    \label{fig:sparsedagfrob500}
  \end{subfigure}
  \caption{ }
  \label{fig:frob500}
\end{figure}

\begin{figure}[H]
  \begin{subfigure}[b]{0.45\textwidth}
    \includegraphics[width=\textwidth]{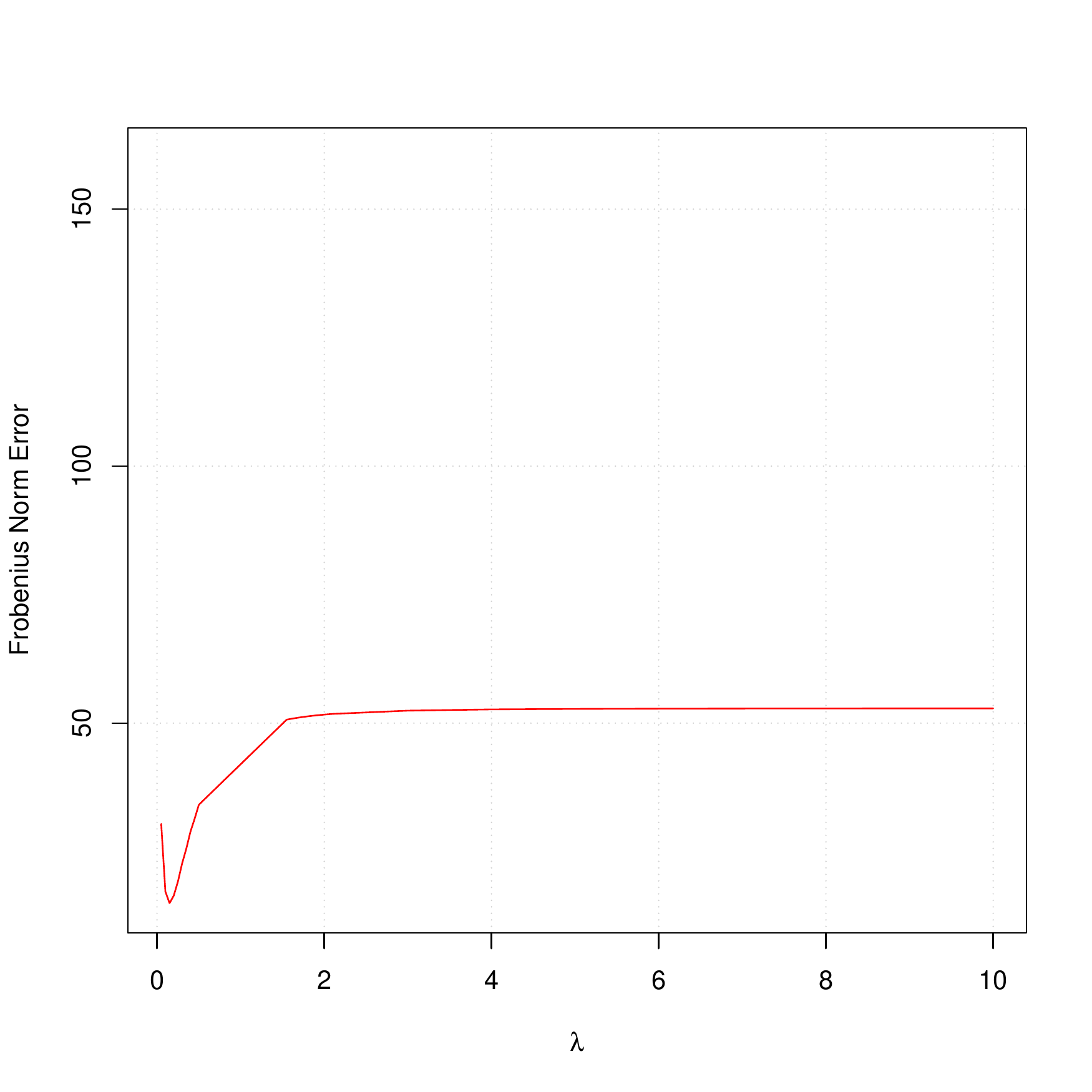}
    \caption{Frobenius Norm Error for CSCS\\
    (y-axis) with varying penalty parameter\\ 
    value (x-axis) for $n = 1500$}
    \label{fig:cscsfrob1500}
  \end{subfigure}
  \begin{subfigure}[b]{0.45\textwidth}
    \includegraphics[width=\textwidth]{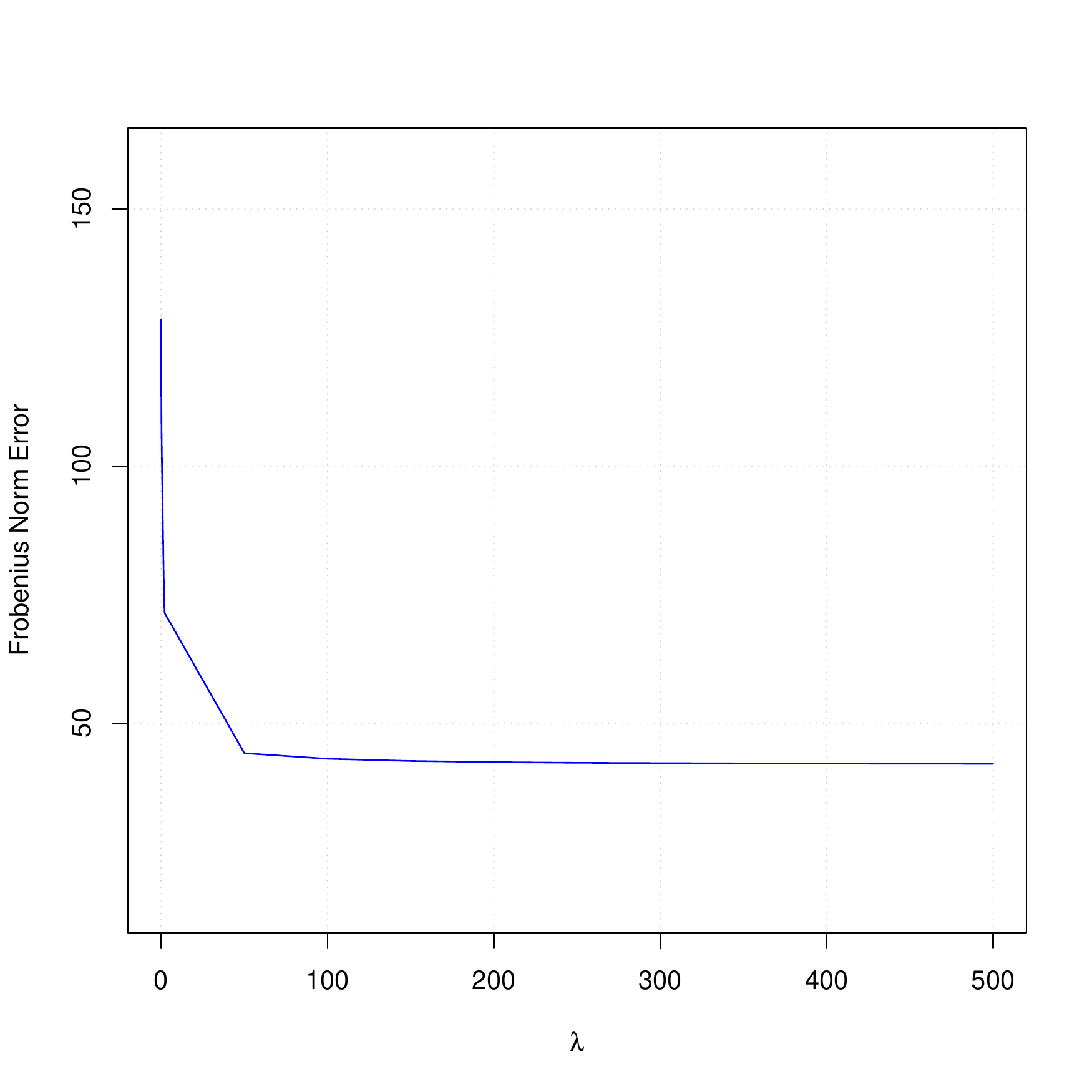}
    \caption{Frobenius Norm Error for 
Sparse DAG averaged over 50 replications for $n = 1500$ for different penalty 
parameter values.}
    \label{fig:sparsedagfrob1500}
  \end{subfigure}
  \caption{ }
  \label{fig:frob1500}
\end{figure}

As mentioned in Section \ref{sec:intro}, the assumption $\{D_{ii}\}_{i=1}^p$ are 
identically $1$ cannot be accounted for/justified by preprocessing the data, and can 
affect the estimation performance of the Sparse DAG approach. To illustrate this fact, 
we consider the settings $p=1000$ and $n=p/2, 3p/2$ and generate $50$ datasets for 
a range of $\lambda$ values similar to the model selection experiment above. Figures
\ref{fig:frob500} and \ref{fig:frob1500} show the Frobenius norm difference (averaged 
over 50 independent repetitions) between the true inverse covariance matrix 
and the estimate ($||\Omega - \hat{\Omega}||_F$), where $\hat{\Omega}$ is the 
estimated inverse covariance matrix for CSCS and Sparse DAG for a range on 
penalty parameter values for $n = 500$ and $n = 1500$ respectively. 

For each method (CSCS and Sparse DAG), we start with a penalty parameter value 
near zero ($0.01$) and increase it till the Frobenius norm error becomes constant, i.e., 
the penalty parameter is large enough so that all the off-diagonal entries of the 
Cholesky parameter are set to zero. That is why the range of penalty parameter 
values for the error curves is different in the (a) and (b) parts of 
Figures \ref{fig:frob500} and \ref{fig:frob1500}. For $n=500$, CSCS achieves a 
minimum error value of $19.9$ at $\lambda = 0.2$, the maximum error value of 
$52.8$ is achieved at $\lambda = 5$ (or higher) when the resulting estimate of 
$\Omega$ is a diagonal matrix with the $i^{th}$ diagonal entry given by $1/S_{ii}$ 
for $1 \leq i \leq p$. On the the other hand, Sparse DAG achieves a minimum error 
value of $42$ at $\lambda = 300$ (or higher) when the resulting estimate of 
$\Omega$ is the identity matrix, and achieves a maximum error value of $121.4$ at 
$\lambda = 0.1$. If the penalty parameter is chosen by BIC (see Table \ref{tab:frob}) 
then CSCS has an error value of $22$ (corresponding to $\lambda = 0.3$) and Sparse 
DAG has an error value of $97$ (corresponding to $\lambda = 0.35$). A similar 
pattern is observed for the case $n = 1500$. It is clear that CSCS has a significantly 
superior overall estimation performance than Sparse DAG in this setting. 
\begin{table}[H]
\resizebox{\textwidth}{!}{
    \centering
    \begin{tabular}{||c||cc||cc||cc||cc||}
      \hline
      & \multicolumn{2}{c||}{$\bf n=125$} 
      & \multicolumn{2}{c||}{$\bf n=250$} 
      & \multicolumn{2}{c||}{$\bf n=500$}
& \multicolumn{2}{c||}{$\bf n=1500$}\\\hline
      Solver  & Mean & Std. Dev.   & Mean & Std. Dev. & Mean & Std. Dev. & Mean & Std. Dev.  \\
      \hline                                 
      Sparse Cholesky  & 0.012796  & 0.000045 & 0.018461
&0.000108 & 0.078832 &0.000122 & 0.127916 & 0.000027\\ 
      Sparse DAG & 0.113955  & 0.000200 & 0.129142
&0.000048 & 0.135271 &0.000066 & 0.138633 & 0.000026\\ 
      CSCS & {\bf 0.118440}  & 0.000111 & {\bf 0.133958} &0.000036
              & {\bf 0.138492} &0.000023 & {\bf 0.139891} & 0.000001\\ 
      \hline
    \end{tabular}
}
    \caption{Mean and Standard Deviation of area-under-the-curve (AUC) for 100 
    simulations for p = 1000. Each simulation yields a ROC curve from which the AUC is 
    computed for FPR in the interval [0.01, 0.15]. CSCS achieves the highest AUC in each column.}
    \label{tbl:aucp1000}
\end{table}

\begin{table}[H]
\resizebox{\textwidth}{!}{
    \centering
    \begin{tabular}{||c||cc||cc||cc||cc||}
      \hline
      & \multicolumn{2}{c||}{$\bf n=250$} 
      & \multicolumn{2}{c||}{$\bf n=500$} 
      & \multicolumn{2}{c||}{$\bf n=1000$}
& \multicolumn{2}{c||}{$\bf n=3000$}\\\hline
      Solver  & Mean & Std. Dev.   & Mean & Std. Dev. & Mean & Std. Dev. & Mean & Std. Dev.  \\
      \hline                                 
      Sparse Cholesky  & 0.015131  & 0.000050 & 0.032391
&0.000105 & 0.124284 &0.000058 & 0.142678 & 0.000012\\ 
      Sparse DAG & 0.141957  &0.000044 & 0.146362
&0.000009 & 0.147984 &0.000005 & 0.148742 & 0.000001\\ 
      CSCS & {\bf 0.144686}  & 0.000019 & {\bf 0.147839} &0.000004
              & {\bf 0.148722} &0.000002 & {\bf 0.148904} & 0.000001\\ 
      \hline
    \end{tabular}
}
    \caption{Mean and Standard Deviation of area-under-the-curve (AUC) for 100 
    simulations for p = 2000. Each simulation yields a ROC curve from which the AUC is 
    computed for FPR in the interval [0.001, 0.15]. CSCS achieves the highest AUC in 
    each column.}
    \label{tbl:aucp2000}
\end{table}

\begin{table}[H]
\centering
\begin{tabular}{||c|c|c||}
\hline
& $n = 500$ & $n = 1500$\\
\hline
  CSCS & {\bf 22.03} (0.09) & {\bf 16.44} (0.06)\\
  Sparse DAG & 96.98(0.81) & 108.90(0.14)\\
   \hline
\end{tabular}
\caption{Frobenius Norm error for $\lambda$ chosen by BIC for CSCS and Sparse 
DAG for $p=1000$}
\label{tab:frob}
\end{table}

\subsection{Application to genetics data} \label{sec:applicationgen}

\begin{figure}[H]
  \centering
    \begin{subfigure}[b]{0.4\textwidth}
        \includegraphics[width=\textwidth]{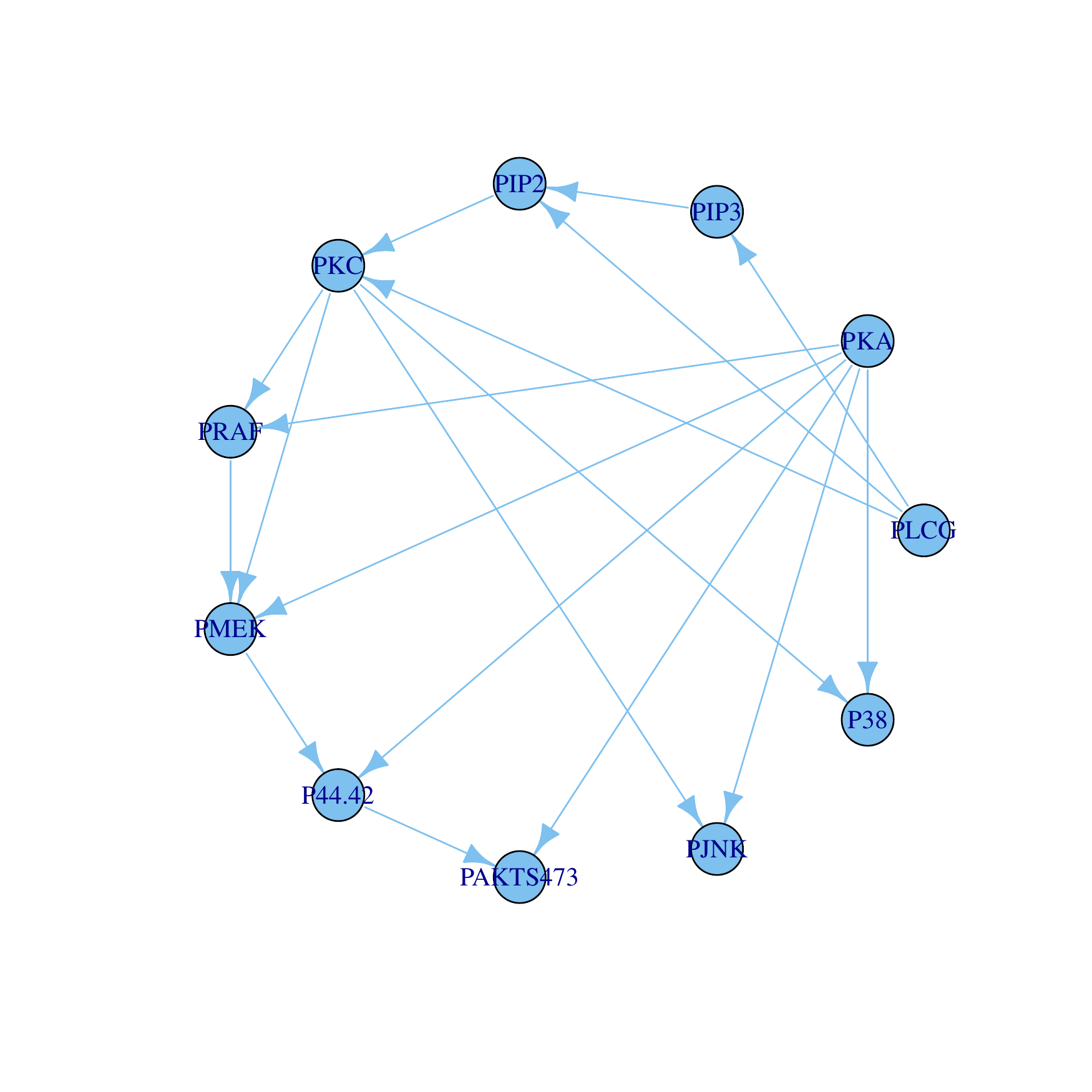}
        \caption{Sachs}
        \label{tg}
    \end{subfigure}
~
    \begin{subfigure}[b]{0.4\textwidth}
        \includegraphics[width=\textwidth]{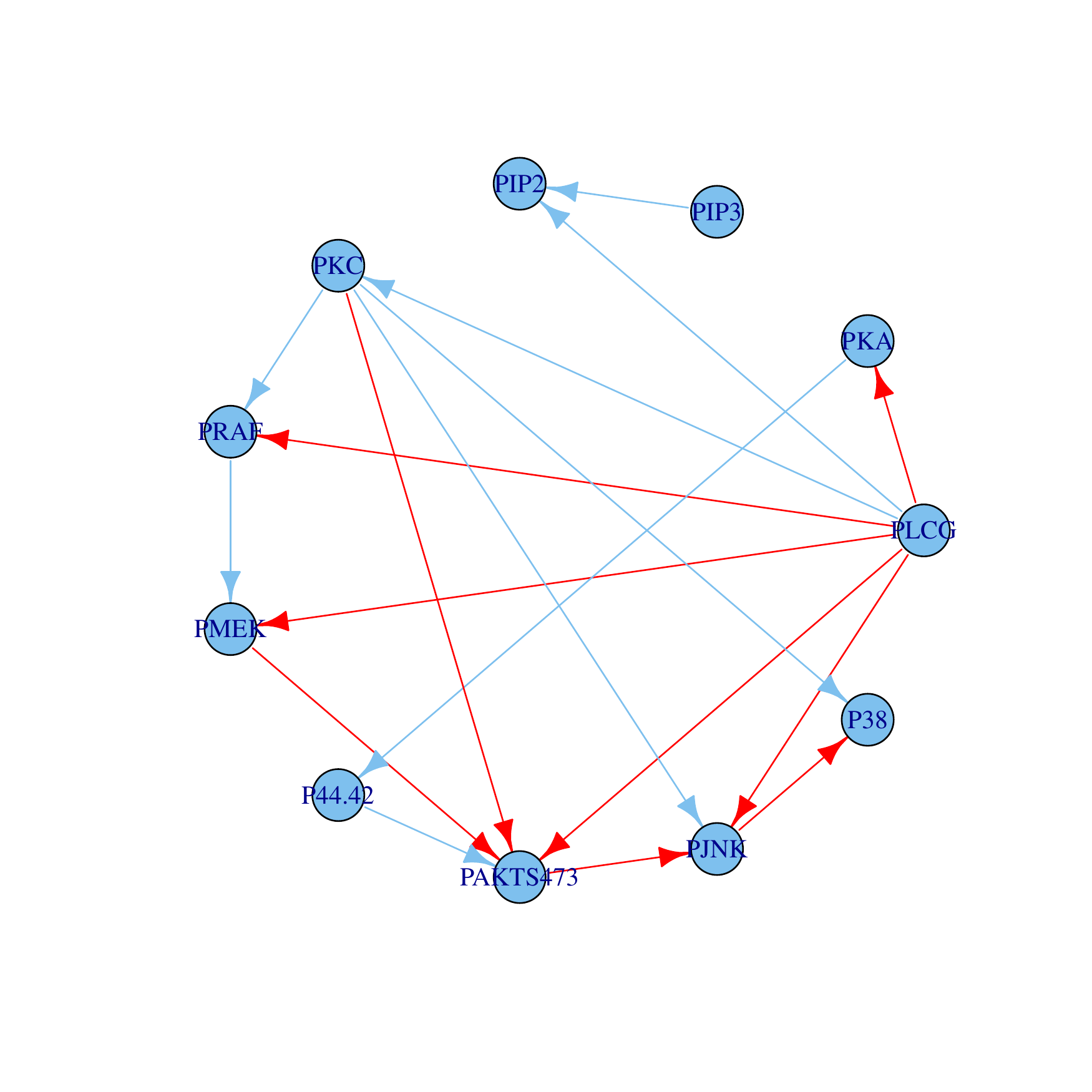}
        \caption{CSCS }
        \label{cg}
    \end{subfigure}
\\
    \begin{subfigure}[b]{0.4\textwidth}
        \includegraphics[width=\textwidth]{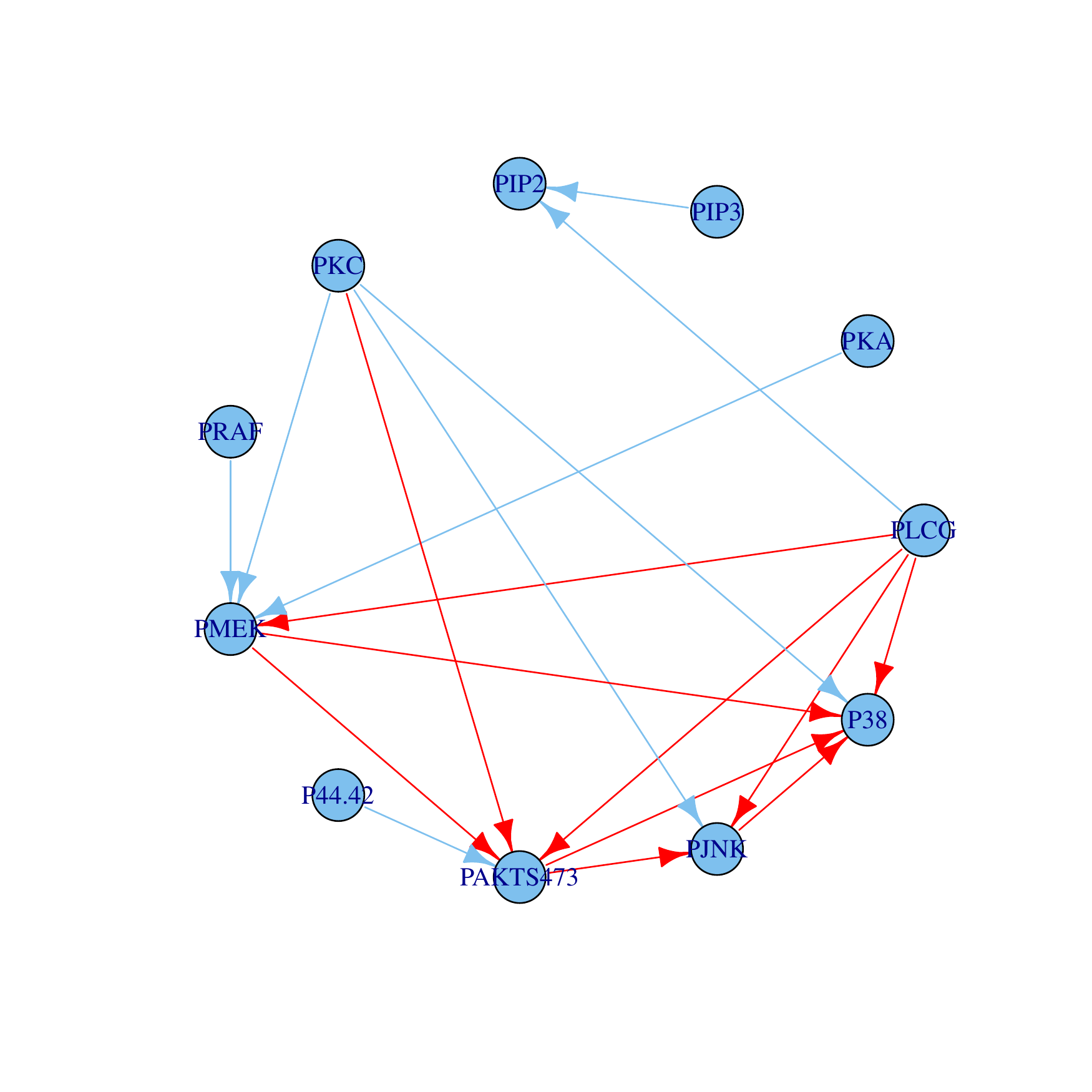}
        \caption{Sparse Cholesky}
        \label{hg}
    \end{subfigure}
~
    \begin{subfigure}[b]{0.4\textwidth}
        \includegraphics[width=\textwidth]{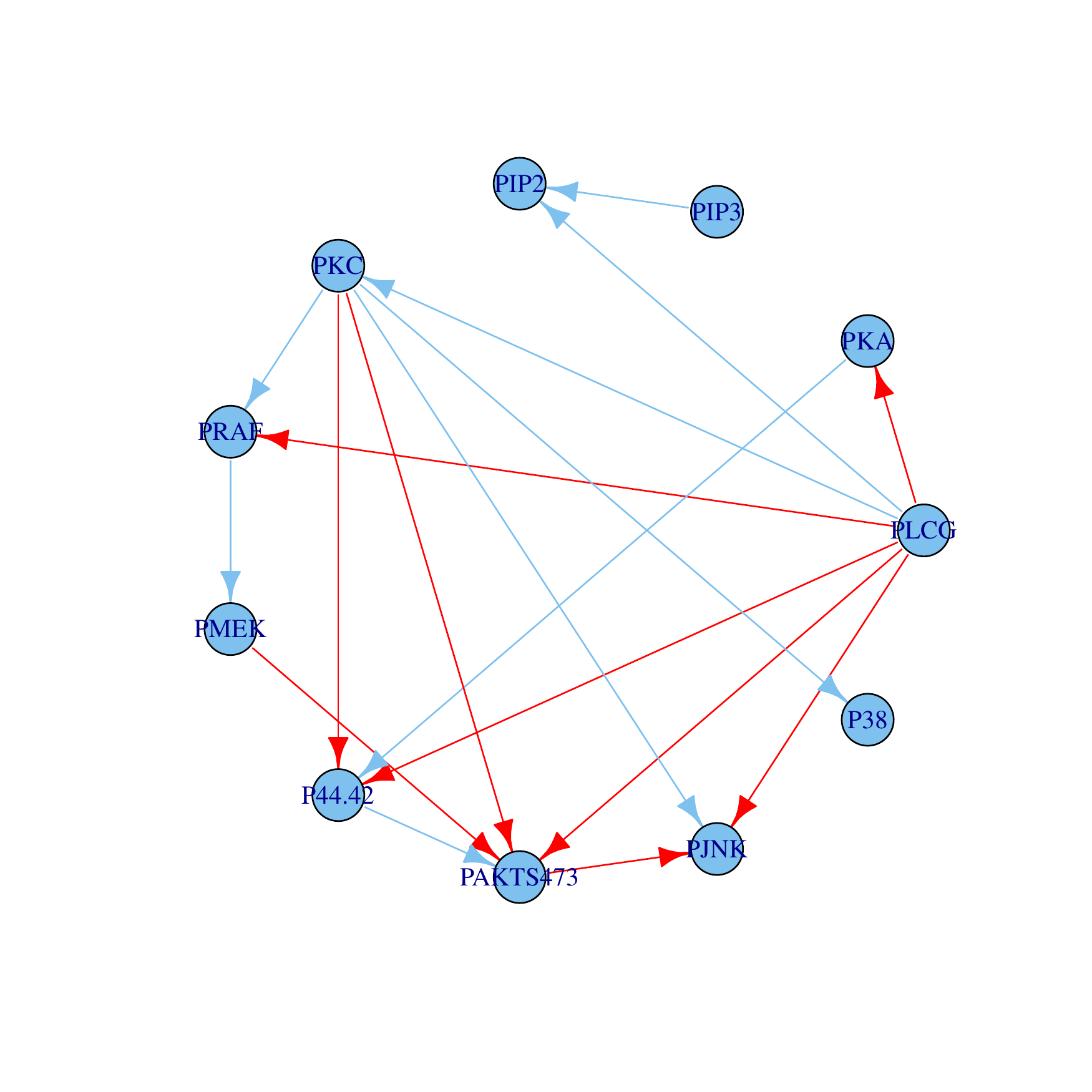}
        \caption{Sparse DAG}
        \label{sg}
    \end{subfigure}
\caption{True and estimated graphs from cell-signaling data. A blue arrow
denotes a true positive, while a red arrow denotes a false positive} 
\label{fig:cellgraphs}
\end{figure}

\begin{figure}[H]
  \centering
    \begin{subfigure}[b]{0.4\textwidth}
        \includegraphics[width=\textwidth]{truegraph.pdf}
        \caption{Sachs}
        \label{fig:tg}
    \end{subfigure}
~
    \begin{subfigure}[b]{0.4\textwidth}
        \includegraphics[width=\textwidth]{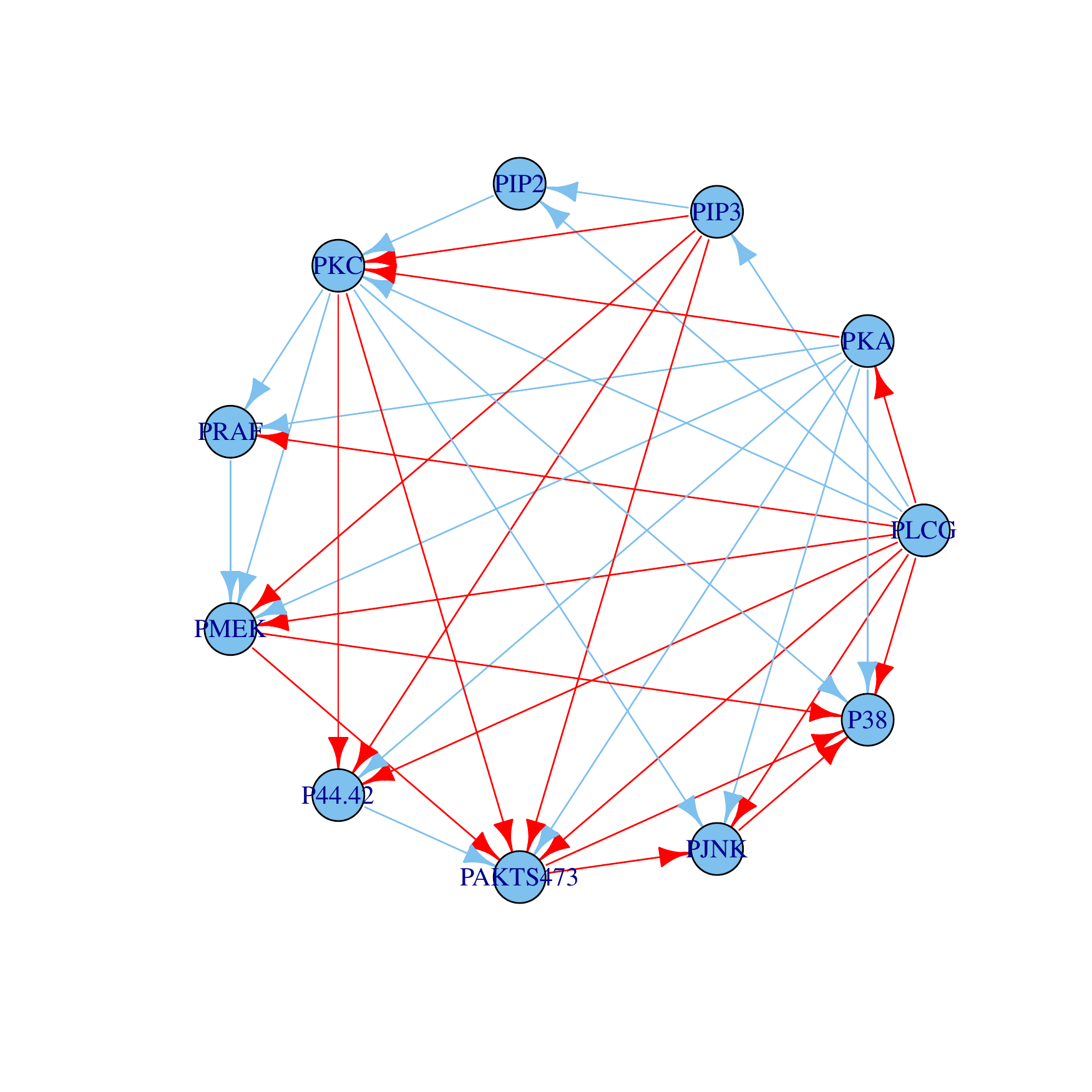}
        \caption{CSCS}
        \label{fig:cg}
    \end{subfigure}
\\
    \begin{subfigure}[b]{0.4\textwidth}
        \includegraphics[width=\textwidth]{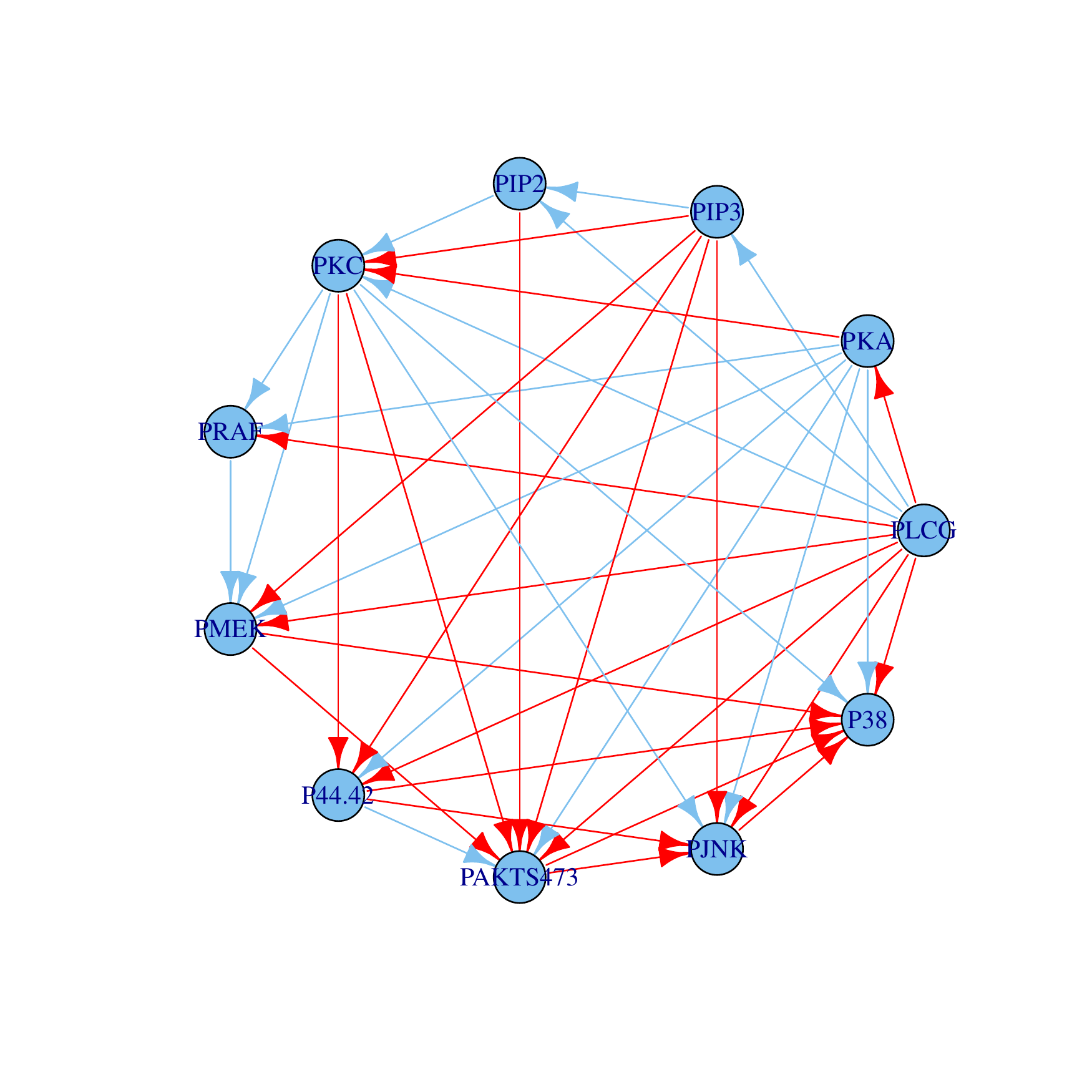}
        \caption{Sparse Cholesky}
        \label{fig:hg}
    \end{subfigure}
~
    \begin{subfigure}[b]{0.4\textwidth}
        \includegraphics[width=\textwidth]{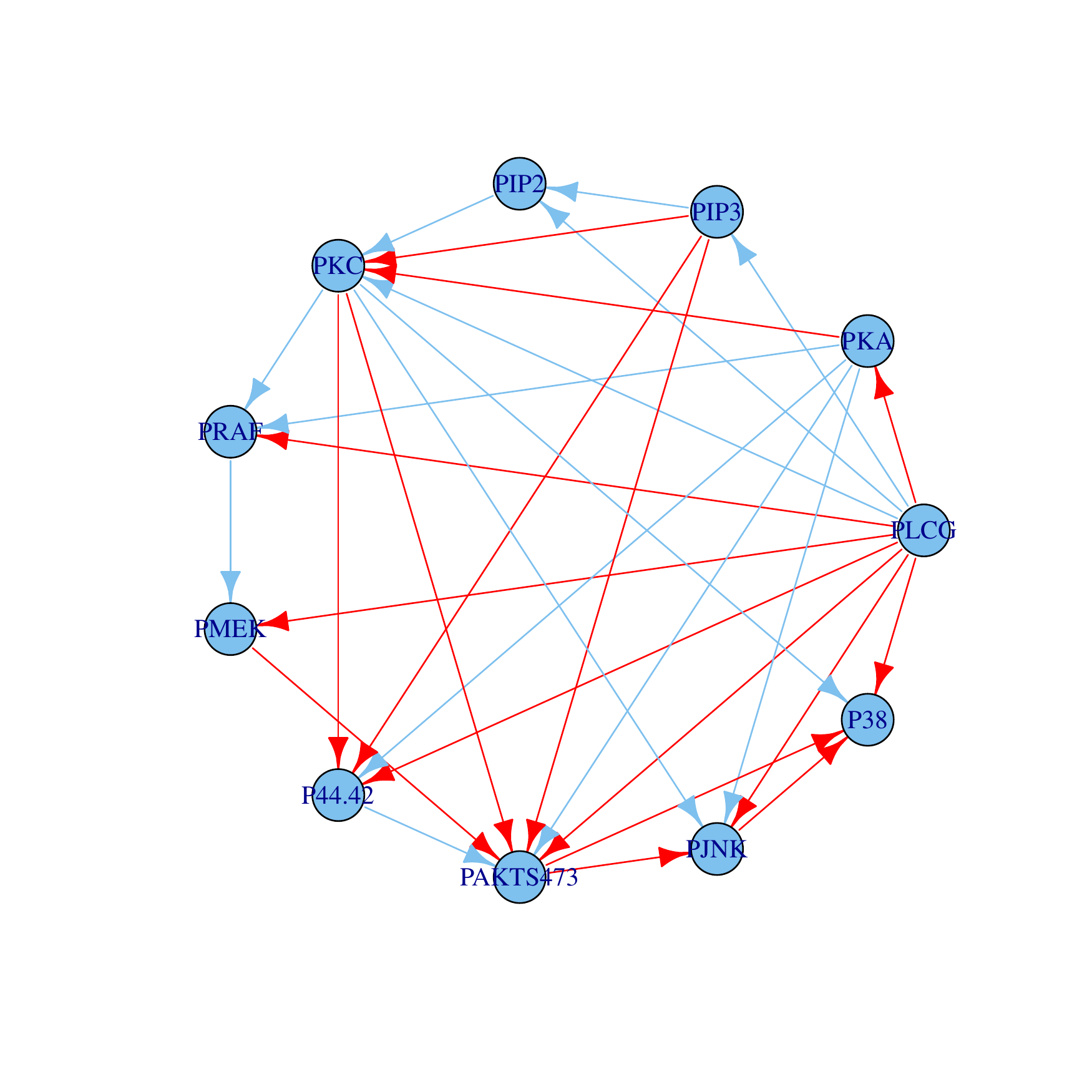}
        \caption{Sparse DAG}
        \label{fig:sg}
    \end{subfigure}
\caption{True and estimated graphs from cell-signaling data by setting
  $\lambda_i(\alpha) = 2 n ^{-\frac{1}{2}}
  Z_{\frac{\alpha}{2p(i-1)}}^*$. A blue arrow
denotes a true positive, while a red arrow denotes a false positive} 
\label{fig:cellgraphscv}
\end{figure}

In this section, we analyze a flow cytometry dataset on p = 11 proteins and n = 7466 
cells, from \cite{Sachs2003}. These authors fit a directed acyclic graph (DAG) to the 
data, producing the network in Figure \ref{fig:tg}. The ordering of the connections 
between pathway components were established based on perturbations in cells using 
molecular interventions and we consider the ordering to be known a priori.This dataset 
is analyzed in \cite{FHT:2008a} and \cite{Shajoie:Michalidis:2010} using the
Glasso algorithm and the Sparse DAG algotirms, respectively. In
\cite{FHT:2008a}, the authors estimated the many graphs by varying
the $\ell_1$ penalty and report around 50\%
false positive and false negative rates between one of the estimates
and the findings of \cite{Sachs2003}. Figure \ref{fig:cellgraphs}
shows the true graph as well as the estimated graph using CSCS, Sparse Cholesky 
and Sparse DAG. We pick the penalty parameter by matching the sparsity to the true 
graph (approximately 72\%). Here both Sparse DAG and CSCS perform better than 
Sparse Cholesky. 

In \cite{Shajoie:Michalidis:2010}, the authors 
recommend using the following equation for penalty parameter selection:
$\lambda_i(\alpha) = 2 n ^{-\frac{1}{2}} Z_{\frac{\alpha}{2p(i-1)}}^*$, where $Z_q^**$ 
denotes the $(1-q)th$ quantile of the standard normal distribution. This choice uses a 
different penalty parameter for each row, and all the three penalized methods 
(Sparse Cholesky, Sparse DAG, CSCS) can be easily adapted to incorporate this. 
Using this method for Sparse DAG gives us a false positive rate of $0.46$ and a true 
positive rate of $0.78$, while Sparse Cholesky has a false positive rate of $0.62$ and 
a true positive rate of $0.94$. Hence, while Sparse Cholesky tends to find a
lot of false edges, it fails to detect only one true edge. CSCS also fails to detect only 
one edge and thus has a true positive rate of $0.94$. However, it does better overall 
as indicated by the lower false positive rate at $0.51$. Figure \ref{fig:cellgraphscv}
shows the true graph as well as the estimated graph using CSCS, Sparse Cholesky 
and Sparse DAG. By picking the penalty parameter according to BIC, Sparse 
Cholesky results in a completely sparse graph while CSCS and Sparse DAG return 
very dense graphs. The true and false positives for the 72\% sparsity, normal 
quantile and BIC based estimates are provided in Table \ref{tab:cellspar}. 
\begin{table}[H]
\resizebox{\textwidth}{!}{
\centering
\begin{tabular}{||c||cc||cc||cc||}
  \hline
&\multicolumn{2}{c||}{72\% Sparsity} 
&\multicolumn{2}{c||}{$\lambda_i(\alpha) = 2 n ^{-\frac{1}{2}}
  Z_{\frac{\alpha}{2p(i-1)}}^*$}
&\multicolumn{2}{c||}{BIC}\\
\hline
  Solver & FP & TP & FP & TP & FP & TP \\ 
  \hline
  CSCS & 0.2432 & 0.5000 & 0.5135 & 0.9444 & 0.8649 & 1.0000 \\ 
    Sparse Cholesky & 0.2703 & 0.4444 & 0.6216 & 0.9444 & 0.0000 & 0.0000\\ 
    Sparse DAG & 0.2432 & 0.5000 & 0.4595 & 0.7778 & 0.8108 & 1.0000\\ 
   \hline
\end{tabular}
}
\caption{TPR \& FPR for Cell Signalling Pathway Data}
\label{tab:cellspar}
\end{table}

\subsection{Application to call center data} \label{sec:applicationcall}

In this section we discuss the application of CSCS, Sparse Cholesky and Sparse DAG 
to the call center data from \cite{HLPL:2006}. The data, coming from one call center in 
a major U.S. northeastern financial organization, contain the information about the 
time every call arrives at the service queue. For each day in 2002, except for 6 days 
when the data-collecting equipment was out of order, phone calls are recorded from 
7:00am until midnight. The 17-hour period is divided into 102 10-minute intervals, and 
the number of calls arriving at the service queue during each interval are counted. 
Since the arrival patterns of weekdays and weekends differ, the focus is on weekdays 
here. In addition, after using singular value decomposition to screen out outliers that 
include holidays and days when the recording equipment was faulty (see 
\cite{SH2005}), we are left with observations for 239 days. 

The data were ordered by time period. Denote the data for day $i$ by $N_i = (N_{i,1}, 
\hdots, N_{i,102})'$, $i = 1, \hdots, 239$ where $N_{i,t}$ is the number of calls arriving 
at the call centre for the $t^{th}$ 10-minute interval on day $i$. Let $y_{it} = \sqrt{N_{it}
+1/4}, i = 1, \hdots, 239, t = 1, \hdots, 102$. We apply the three penalized likelihood 
methods (CSCS, Sparse DAG, Sparse Cholesky) to estimate the $102 \times 102$ 
covariance matrix based on the residuals from a fit of the saturated mean model. 
Following the analysis in \cite{HLPL:2006}, the $\ell_1$ penalty parameter for all three 
methods was picked using $5$-fold cross validation on the training data set as follows. 
Randomly split the full dataset $D$ into $K$ subsets of about the same size, denoted 
by $D_v, v = 1,...,K$. For each $v$, we use the data $D - D_v$ to estimate 
$\Sigma_{-v}$ and $D_v$ to validate. Then pick $\lambda$ to minimize:
$$
\text{CV}(\lambda) = \frac{1}{K} \sum_{v = 1}^K \big(d_v \log
|{\hat{\Sigma}_{-v}}| + \sum_{i \in I_v} y_i' \hat{\Sigma}_{-v}^{-1}
y_i \big)
$$

\noindent
where $I_v$ is the index set of the data in $D_v$, $d_v$ is the size of $I_v$, 
and $\hat{\Sigma}_v$ is the variance-covariance matrix estimated using the 
training data set $D - D_v$. 

To assess the performance of different methods, we split the 239 days into training 
and test datasets. The data from the first $T$ days ($T = 205, 150, 100, 75$), form the 
training dataset that is used to estimate the mean vector and the covariance matrix. 
The mean vector is estimated by the mean of the training data vectors. Four different 
methods, namely, CSCS, Sparse Cholesky, Sparse DAG and S (sample covariance 
matrix) are used to get an estimate of the covariance matrix. For each of the three 
penalized methods, the penalty parameter is chosen both by cross-validation and the 
BIC criterion. Hence, we have a total of seven estimators for the covariance matrix. 
The log-likelihood for the test dataset (consisting of the remaining $239-T$ days) 
evaluated at all the above estimators is provided in Table \ref{likelihood}. For all 
training data sizes, CSCS clearly demonstrates superior performance as compared to 
the other methods. Also, the comparative performance of CSCS with other methods 
improves significantly with decreasing training data size. 
\begin{table}[H]
\centering
\begin{tabular}{||c|c|c|c|c||}
  \hline
\hline
& \multicolumn{4}{c||}{Training Data Size} \\
\hline
Method & $205$ & $150$ & $100$ & $75$ \\ 
  \hline
  CSCS-CV &-1090.447  & -1369.181 & -2225.907&\bf{-2841.348}\\ 
  CSCS-BIC &\bf{-1072.75}&\bf{ -1364.145} &\bf{-2214.729}&-2849.931\\ 
  Sparse DAG-CV &-1077.791&-2237.298 &-3576.343& -4499.298\\ 
  Sparse DAG-BIC &-1135.980 &-2421.950 &-3817.689&-4846.118\\ 
  Sparse Cholesky-CV &-1500.094 &  -2121.005&-3579.932&-496617558322\\ 
  Sparse Cholesky-BIC &-1523.409& -2178.738&-3584.160 &-5444.471\\ 
  S& -1488.224& -7696.740& not pd&not pd\\ 
   \hline
\hline
\end{tabular}
\caption{Test data log-likelihood values for various estimation methods with 
training data size $205, 150, 100, 75$. The maximum likelihood value in each column 
is written in bold.}
\label{likelihood}
\end{table}

Huang et al. \cite{HLPL:2006} additionally use the estimated mean and covariance 
matrix to forecast the number of arrivals in the later half of the day using arrival 
patterns in the earlier half of the day. Following their method, we compared the 
performance of all the four methods under consideration (details provided in 
Supplemental Section \ref{ccforecast}). We found that all the three penalized 
methods outperform the sample covariance matrix estimator. However, as far as this 
specific forecasting task is concerned, the differences in their performance compared 
to each other are marginal. We suspect that the for the purposes of this 
forecasting task, the estimated mean (same for all three methods) has a much 
stronger effect than the estimated covariance matrix. Hence the difference in  
forecasting performance is much smaller than the difference in likelihood values.  
Nevertheless, Sparse Cholesky has the best performance for training data size $T 
= 205, 150$ (when the sample size is more than the number of variables) and 
CSCS has the best performance for training data sizes $T = 100, 75$ (when the 
sample size is less than the number of variables). See 
Supplemental Section \ref{ccforecast} for more details.


\section{Asymptotic properties} \label{sec:asymptotics}

\noindent
In this section, asymptotic properties of the CSCS algorithm will be examined in a 
high-dimensional setting, where the dimension $p = p_n$ and the penalty parameter 
$\lambda = \lambda_n$ vary with $n$. In particular, we will establish estimation 
consistency and model selection consistency (oracle properties) for the CSCS 
algorithm under suitable regularity assumptions. Our approach is based on the 
strategy outlined in Meinshausen and Buhlmann \cite{Meinshausen:Buhlmann:2006} 
and Massam, Paul and Rajaratnam \cite{MPR:2007}. A similar approach 
was used by Peng et al. \cite{PWZZ:2009} to establish asymptotic properties of 
SPACE, which is a penalized pseudo likelihood based algorithm for sparse estimation 
of $\Omega$. Despite the similarity in the basic line of attack, there is an important 
structural difference between the asymptotic consistency arguments in 
\cite{PWZZ:2009} and this section (apart from the fact that we are imposing sparsity 
in $L$, not $\Omega$). For the purpose of proving asymptotic consistency, the authors 
in \cite{PWZZ:2009} assume that diagonal entries of $\Omega$ are known, thereby 
reducing their objective function to the sum of a quadratic term and an $\ell_1$ penalty 
term in $\Omega$. The authors in \cite{Shajoie:Michalidis:2010} also establish graph 
selection consistency of the Sparse DAG approach under the assumption that the 
diagonal entries of $L$ are $1$. We do not make such an assumption for $L$, which 
leaves us with $p$ additional non-zero parameters, and additional logarithmic terms in 
the objective function to work with. Nevertheless, we are able to adapt the basic 
consistency argument in this challenging setting with an almost identical set of 
regularity assumptions as in \cite{PWZZ:2009} (with assumptions on $\Omega$ 
replaced by the same assumptions on $L$). In particular, we only replace two 
assumptions in \cite{PWZZ:2009} with a weaker and a stronger version respectively 
(see Assumption (A4) and  Assumption (A5) below for more details)


We start by establishing the required notation.  Let $\{\bar{\Omega}_n = \bar{L}_n^t \bar{L}_n\}_{n \geq 1}$ 
denote the sequence of true inverse covariance matrices, and $\bar{\boldsymbol \eta}_n^r$ denote the 
lower triangular entries (including the diagonal) in the $r^{th}$ row of $\bar{L}_n$, for $1 \leq r \leq p$. Let 
$\mathcal{A}_n^r$ denote the set of indices corresponding to non-zero entries in $r^{th}$ row of $\bar{L}_n$ 
for $1 \leq r \leq p$, and let $q_n =  \sum_{r=1}^{p_n} \left| \mathcal{A}_n^r \right|$. Let $\bar{\Sigma}_n = 
\bar{\Omega}_n^{-1}$ denote the true covariance matrix for every $n \geq 1$. The following standard 
assumptions are required. 
\begin{itemize}
\item (A1 - Bounded eigenvalues) The eigenvalues of $\bar{\Omega}_n$
  are bounded below by $\lambda_{min} > 0$, and bounded above by
  $\lambda_{max} < \infty$ uniformly for all $n$.
\item (A2 - Sub Gaussianity) The random vectors ${\bf Y}^1,\dots,{\bf
    Y}^n$ are \emph{i.i.d.}  sub-Gaussian for every $n \geq 1$, i.e.,
  there exists a constant $c > 0$ such that for every ${\bf x} \in
  \mathbb{R}^{p_n}$, $E \left[ e^{{\bf x}' {\bf Y}^i} \right] \leq
  e^{c {\bf x}' {\bf x}}$. 
\item (A3 - Incoherence condition) There exists $\delta < 1$ such that
  for every $n \geq 1$, $1 \leq r \leq p_n$ and $j \notin \mathcal{A}_n^r$, 
$$
\left| \bar{\Sigma}_{n,j,\mathcal{A}^r}^t \left( \bar{\Sigma}_{n, \mathcal{A}^r \mathcal{A}^r} + 
\frac{2}{(\bar{\eta}_r^r)^2} \Delta_r \right)^{-1} \mbox{sign} \left( \bar{\boldsymbol 
\eta}_{\mathcal{A}^r}^r \right) \right| \leq \delta. 
$$

\noindent
Here, $\Delta_r$ is a $|\mathcal{A}^r| \times |\mathcal{A}^r|$ matrix with 
$$
(\Delta_r)_{jj'} = \begin{cases} 
1 & \mbox{if } j = j' = |\mathcal{A}^r|, \cr 
0 & \mbox{otherwise}. 
\end{cases}
$$

\item (A4 - Signal size) For every $n \geq 1$, let 
$$
s_n = \min_{1 \leq r \leq p} \min_{j \in \mathcal{A}_n^r} \left| \bar{\eta}^r_{n,j} \right|. 
$$

\noindent
Then $\frac{s_n}{\sqrt{d_n} \lambda_n} \rightarrow \infty$, where $d_n = \max_{1 \leq r \leq p_n} 
|\mathcal{A}^r|$. This assumption will be useful for establishing sign consistency. The signal size 
condition in \cite{PWZZ:2009} is $\frac{s_n}{\sqrt{q_n} \lambda_n} \rightarrow \infty$, which is 
stronger than the signal size condition above, as $d_n \leq q_n$. 
\item (A5 - Growth of $p_n$, $q_n$ and $\lambda_n$) The following conditions hold: 
$p_n = O(n^\kappa)$ for $\kappa \geq 0$, $q_n = o \left( \sqrt{\frac{n}{\log n}} \right)$, $\sqrt{\frac{q_n 
\log n}{n}} = o(\lambda_n)$,  $\lambda_n \sqrt{\frac{n}{\log n}} \rightarrow \infty$ and $q_n \lambda_n 
\rightarrow 0$ as $n \rightarrow \infty$. The growth conditions in \cite{PWZZ:2009} are the same as above 
(with $q_n$ denoting the sparsity in the true $\Omega$ in \cite{PWZZ:2009}), expect that $q_n \lambda_n 
\rightarrow 0$ above is replaced by the weaker assumption $\sqrt{q_n} \lambda_n \rightarrow 0$. 
\end{itemize}

\noindent
Under these assumptions, the following consistency result can be established. 
\begin{thm} \label{thm:consistency}
Suppose that (A1)-(A5) are satisfied. Then there exists a constant $C > 0$, such that for any $\gamma > 0$, 
the following events hold with probability at least  $1 - O(n^{-\eta})$: 
\begin{enumerate}[(i)]
\item A solution of the minimization problem 
\begin{equation} \label{eq14}
\inf_{L \in \mathcal{L}_{p_n}} Q_{CSCS} (L) 
\end{equation}

\noindent
exists. 
\item (Estimation and sign consistency): any solution $\hat{L}_n$ of the minimization problem in (\ref{eq14}) 
satisfies 
$$
\|\hat{L}_n - \bar{L}_n\| \leq C q_n \lambda_n. 
$$

\noindent
and 
$$
\mbox{sign}(\hat{L}_{n,ij}) = \mbox{sign}(\bar{L}_{n,ij}), 
$$

\noindent
for every $1 \leq j \leq i \leq p$. 
\end{enumerate}
\end{thm}

\noindent
Here $\mbox{sign}(x)$ takes the values $\{-1,0,1\}$ when $x < 0$, $x = 0$, and $x > 
0$ respectively. A proof of the above result is provided in the appendix.

\section{Discussion} \label{sec:discussion}

\noindent
This paper proposes a novel penalized likelihood based approach for estimation and 
model selection in Gaussian DAG models. The goal is to overcome some of the 
shortcomings of current methods, but at the same time retain their respective 
strengths. We start with the objective function for the highly useful Sparse Cholesky 
approach in \cite{HLPL:2006}. Reparametrization of this objective function in terms of 
the inverse of the classical Cholesky factor of the covariance matrix, along with 
appropriate changes to the penalty term, leads us to the formulation of the CSCS 
objective function. It is then shown that the CSCS objective function is jointly convex in 
its arguments. A coordinate-wise minimization algorithm that minimizes this objective, 
via closed form iterates, is proposed, and subsequently analyzed. The convergence of 
this coordinate-wise minimization algorithm to a global minimum is established 
rigorously.   It is also established that the estimate produced by the CSCS algorithm 
always leads to a positive definite estimate of the covariance matrix - thus ensuring 
that CSCS leads to well defined estimates that are always computable. Such a 
guarantee is not available with the Sparse Cholesky approach when $n < p$. Large 
sample properties of CSCS establish estimation and model selection consistency of 
the method as both the sample size and dimension tend to infinity. We also point out 
that the Sparse DAG approach in \cite{Shajoie:Michalidis:2010}, while always useful 
for graph selection, may suffer for estimation purposes due the assumption that the 
conditional variances $\{D_{ii}\}_{i=1}^p$ are identically $1$. The performance of 
CSCS compared to Sparse Cholesky and Sparse DAG is also illustrated via 
simulations and application to a cell-signaling pathway dataset and a call center 
dataset. These experiments complement and support the technical results in the 
paper by demonstrating the following. 
\begin{enumerate}[(a)]
\item When $n < p$, it is easy to find examples where Sparse Cholesky converges to 
its global minimum which corresponds to a singular covariance matrix 
(Section \ref{sec:sparsecholconv}). 
\item When $n < p$, the graph selection and estimation performance of CSCS is 
significantly better than Sparse Cholesky, due to the fact that Sparse Cholesky either 
converges to a global minimum with singularity issues, or to a local minimum 
(Section \ref{sec:simulated} and Section \ref{sec:applicationcall}). 
\item For graph selection, CSCS is competitive with Sparse DAG and can have better 
performance as compared to Sparse DAG. Although the improvement may not 
sometimes be as significant as that over Sparse Cholesky, these results demonstrate 
that CSCS is a useful addition to the high-dimensional DAG selection toolbox 
(Section \ref{sec:simulated} and Section \ref{sec:applicationgen}).
\item For estimation purposes, CSCS can lead to significant improvements in 
performance over Sparse DAG (Section \ref{sec:simulated}). 
\end{enumerate}

\renewcommand{\thesection}{\Alph{section}}
\setcounter{section}{0}

\renewcommand{\thelemma}{$\mathcal{S}$.\arabic{lemma}}
\setcounter{lemma}{0}

\renewcommand{\theequation}{$\mathcal{S}$.\arabic{equation}}
\setcounter{equation}{0}

\bibliographystyle{plain}
\bibliography{references}

\newpage
\pagenumbering{gobble}
\pagenumbering{arabic}

\section*{Supplemental Document for ``A convex framework for high-dimensional 
sparse Cholesky based covariance estimation"}

\section{Proof of Lemma \ref{leminfimum}} 

\noindent
Note that $S_n$ (an $n \times n$ sample covariance matrix for the first $n$ variables) 
is non-singular with probaiblity $1$, while $S_{n+1}$ (an $(n+1) \times (n+1)$ matrix) 
is singular with probability $1$. Since 
$$
S_{n+1} = \left[ \begin{matrix} 
S_n & S_{\cdot (n+1)} \cr 
S_{\cdot (n+1)}^t & S_{n+1,n+1} 
\end{matrix} 
\right], 
$$

\noindent
is positive semi-definite, it follows that $S_{n+1,n+1} = S_{\cdot (n+1)}^t S_n^{-1} S_{\cdot (n+1)}$. Hence, 
if ${\boldsymbol \phi}_{n+1}^* = S_n^{-1} S_{\cdot (n+1)}$, we get that 
$$
({\boldsymbol \phi}_{n+1}^*)^t S_n {\boldsymbol \phi}_{n+1}^* + 2 
({\boldsymbol \phi}_{n+1}^*)^t S_{\cdot (n+1)} + S_{n+1,n+1} = 0. 
$$

\noindent
It follows that 
$$
Q_{Chol,n+1} ({\boldsymbol \phi}_{n+1}^*, \frac{1}{m}) = -\log m + \lambda 
\|{\boldsymbol \phi}_{n+1}^*\|_1 \rightarrow -\infty 
$$

\noindent
as $m \rightarrow \infty$. \hfill$\Box$ 

\bigskip

\section{Proof of Lemma \ref{closedform}} 

\noindent
Note that for $1 \leq j \leq k-1$, 
$$
h_{k,A,\lambda} ({\bf x}) = x_j^2 A_{jj} + 2 x_j \left( \sum_{l \neq j} A_{lj} x_l \right) + \lambda |x_j| + 
\mbox{ terms independent of } x_j. 
$$

\noindent
It follows that 
$$
\left( T_j ({\bf x}) \right)_j = \frac{S_{\lambda} \left(- 2 \sum_{l \neq j} A_{lj} x_l \right)}{2 A_{jj}}. 
$$

\noindent
Also, 
$$
h_{k,A,\lambda} ({\bf x}) = -2 \log x_k + x_k^2 A_{kk} + 2 x_k \left( \sum_{l \neq k} A_{lk} x_l \right) 
+ \mbox{ terms independent of } x_k. 
$$

\noindent
It follows that 
\begin{eqnarray*}
  \frac{\partial}{\partial x_k} h_{k,A,\lambda} ({\bf x}) = 0 
  &\Leftrightarrow& -\frac{2}{x_k} + 2 x_k A_{kk} + 2 \sum_{l \neq k} A_{lk} x_l = 0\\
  &\Leftrightarrow& x_k = \frac{- \sum_{l \neq k} A_{lk} x_l  + \sqrt{\left( 
        \sum_{l \neq k} A_{lk} x_l \right)^2 + 4 A_{kk}}}{2 A_{kk}}, 
\end{eqnarray*}

\noindent
Note that since $x_k > 0$ the positive root has been retained as the solution. \hfill$\Box$ 

\bigskip

\section{Proof of Lemma \ref{cmxbasic}}

\noindent
We consider two cases. 

\medskip

\noindent
{\it Case 1 ($n \geq k$)}: It follows from (\ref{eq12}) and (\ref{eq13}) that the update for each 
of the $k$ coordinates in an iteration of Algorithm \ref{algorithm1} can be achieved in $O(k)$ computations. 
Hence, a computational complexity of $O(k^2)$ can be achieved in this case. 

\medskip

\noindent
{\it Case 2 ($n < k$)}: For this case, we will use ideas similar to the analysis of 
computational complexity in [1, 3, 4] in the context of algorithms inducing sparsity 
in $\Omega$. Let ${\bf r} ({\bf x}) = B^T {\bf x} \in \mathbb{R}^n$. Given the initial 
value $\hat{\bf x}^{(0)}$, we evaluate $r(\hat{\bf x}^{(0)}) = B^T \hat{\bf x}^{(0)}$ 
(which takes $O(nk)$) iterations), and keep track of $B^T \hat{\bf x}^{\cur}$ 
throughout the course of the algorithm. Note that if ${\bf x}$ and $\tilde{\bf x}$ 
differ only in one coordinate (say the $m^{th}$ coordinate), then 
$$
(B^T \tilde{\bf x})_j = \sum_{l=1}^k B_{lj} \tilde{x}_l = \sum_{l=1}^k B_{lj} x_l + B_{mj} 
(\tilde{x}_m - x_m) 
$$

\noindent
for every $1 \leq j \leq k$. It follows that it takes $O(n)$ computations to update $B^T {\bf x}$ to 
$B^T \tilde{\bf x}$. Hence, after each coordinatewise update in Algorithm \ref{algorithm1}, it will 
take $O(n)$ computations to update ${\bf r}$ to its current value. For every $1 \leq j \leq k$, note 
that 
$$
\sum_{l \neq j} A_{lj} x_l = \sum_{l=1}^k A_{lj} x_l - A_{jj} x_j = B_{j \cdot} {\bf r} ({\bf x}) - A_{jj} x_j. 
$$

\noindent
where $B_{j \cdot}$ denotes the $j^{th}$ row of the $k \times n$ matrix $B$. It now follows from 
(\ref{eq12}) and (\ref{eq13}) that each coordinatewise update in Algorithm \ref{algorithm1} can 
be performed in $O(n)$ steps. Hence, the computational complexity of $O(nk)$ can be achieved 
for one iteration (which involves $k$ coordinatewise updates) of the Algorithm \ref{algorithm1}. 
\hfill$\Box$ 

\bigskip

\section{Proof of Theorem \ref{thm:convergence}}

\noindent
Fix $1 \leq i \leq p$ arbitrarily. Note that $S_i = \mathbb{Y}_i^T \mathbb{Y}_i$, where 
$\mathbb{Y}_i$ is an $n \times i$ matrix of observations corresponding to the first $i$ 
variables. Since all diagonal entries of $S$ are assumed to be positive, it follows that 
$\mathbb{Y}_i$ has no zero columns. Now, let $\xi \in \mathbb{R}$ be arbitrarily fixed. 
if $Q_{CSCS,i} ({\boldsymbol \eta}^i) < \xi$, then it follows that $ -2 \log \eta^i_i < \xi$ 
(since the other two terms in the expression for $Q_{CSCS,i}$ are non-negative). 
In particular, we obtain that $\eta^i_i > exp(-\xi/2)$. Also, it follows from (\ref{eq9}) that 
$|\eta^i_j| \leq 2 \xi/\lambda$ for every $1 \leq j \leq i-1$, and $\eta^i_i  \leq (\xi + 
b_i)/a_i$. The above arguments, along with the expression for $Q_{CSCS,i}$ in 
(\ref{eq7}) and (\ref{eq8}), and [2, Theorem 2.2] imply that the cyclic coordinatewise 
algorithm for $Q_{CSCS,i}$ will converge to a global minimum of $Q_{CSCS,i}$. It 
follows that Algorithm \ref{algorithm2} converges to a global minimum of 
$Q_{CSCS}$. \hfill$\Box$

\section{Proof of Theorem \ref{thm:consistency}}

\noindent
Note that by (\ref{eq6}), the problem of minimizing $Q_{CSCS}$ with respect to $L$ is equivalent to the 
problem of minimizing $Q_{CSCS, r}$ with respect to ${\boldsymbol \eta}^r$ for $1 \leq r \leq p$. We will 
first establish appropriate consistency results for the minimizers of $Q_{CSCS, r}$, for each $1 \leq r \leq 
p$, and then combine these results to establish Theorem \ref{thm:consistency}. Throughout this proof, we 
will often suppress the dependence of various quantities on $n$, for notational simplicity and ease of 
exposition. We now establish a series of lemmas which will be quite useful in the main proof. 
\begin{lemma} \label{lem:consistency:1}
For any $\gamma > 0$, there exists a constant $C_{\gamma} > 0$ such that with probability at least $1 - 
O(n^{-\gamma})$
$$
\max_{1 \leq i,j, \leq p_n} |S_{ij} - \bar{\Sigma}_{n,ij}| \leq C_{\gamma} \sqrt{\frac{\log n}{n}}.  
$$

\noindent
for large enough $n$. 
\end{lemma}

\noindent
{\it Proof}:  Fix $1 \leq i,j \leq p_n$. Let $\mu_+ := E_{\bar{\Sigma}_n} \left[ (Y^1_i + Y^1_j)^2 \right]$ and 
$\mu_- :=  E_{\bar{\Sigma}_n} \left[ (Y^1_i - Y^1_j)^2 \right]$. It follows that 
\begin{eqnarray}
& & P(|S_{ij} - \bar{\Sigma}_{n,ij}| > t) \nonumber\\
&=& P \left( \left| \frac{1}{n} \sum_{\ell=1}^n (Y^\ell_i + Y^\ell_j)^2 - (Y^\ell_i - Y^\ell_j)^2 - (\mu_+ - \mu_-) 
\right| > 4t \right) \nonumber\\
&\leq& P \left( \left| \frac{1}{n} \sum_{\ell=1}^n (Y^\ell_i + Y^\ell_j)^2 - \mu_+ \right| > 2t \right) + 
P \left( \left| \frac{1}{n} \sum_{\ell=1}^n (Y^\ell_i - Y^\ell_j)^2 - \mu_- \right| > 2t \right). \label{eq18pre1} 
\end{eqnarray}

\noindent
Note that $Y^\ell_i + Y^\ell_j$ are sub-Gaussian random variables 
(by Assumption (A2)) and their variances are uniformly bounded in $i$, $j$ and $n$ 
(by Assumption (A1)). For any $c_1 > 0$, it follows by (\ref{eq18pre1}) and 
[5, Theorem 1.1], that there exist constants $K_1$ and $K_2$ independent of $i$, 
$j$ and $n$ such that 
$$
P \left( |S_{ij} - \bar{\Sigma}_{n,ij}| > C \sqrt{\frac{\log n}{n}} \right) \leq K_1 e^{-K_2 
n \left( c_1 \sqrt{\frac{\log n}{n}} \right)^2} = K_1 e^{-K_2 C^2 \log n} 
$$

\noindent
for large enough $n$. Using the union bound and the fact that $p = O(n^\kappa)$ for 
some $\kappa \geq 0$ gives us the required result. \hfill$\Box$ 

\begin{lemma} \label{lem:consistency:2}
For every $1 \leq r \leq p$, we note that ${\boldsymbol \eta}^r$ minimizes $Q_{CSCS,r}$ 
if and only if 
\begin{eqnarray}
& & d_i^r ({\boldsymbol \eta}^r) = - \lambda_n \mbox{sign}(\eta^r_i) \mbox{ if } \eta^r_i \neq 0, \; 1 \leq i \leq r-1, 
\label{eq15}\\
& & |d_i^r ({\boldsymbol \eta}^r)| \leq \lambda_n \mbox{ if } \eta^r_i = 0, 1 \leq i \leq r-1, \label{eq16}\\
& & d_r^r ({\boldsymbol \eta}^r) = 0, \label{eq17}
\end{eqnarray}

\noindent
where 
\begin{equation} \label{eq18}
d_i^r ({\boldsymbol \eta}^r) = 2 \sum_{j=1}^r \eta^r_j S_{ij} 
\end{equation}

\noindent
for $1 \leq i \leq r-1$, and 
\begin{equation} \label{eq19}
d_r^r ({\boldsymbol \eta}^r) = 2 \sum_{j=1}^r \eta^r_j S_{rj} - \frac{2}{\eta^r_r}. 
\end{equation}

\noindent
Also, if $|d_i^r (\hat{\boldsymbol \eta}^r)| < \lambda_n$ for any minimizer $\hat{\boldsymbol \eta}^r$, then by 
the continuity of $d_i^r$, and the convexity of $Q_{CSCS,r}$, it follows that $\tilde{\eta}_i^r = 0$ for every 
minimizer $\tilde{\boldsymbol \eta}^r$ of $Q_{CSCS,r}$. 
\end{lemma}

\noindent
The proof immediately follows from the KKT conditions for the convex function $Q_{CSCS,r}$. 
\begin{lemma} \label{lem:consistency:3}
For every $1 \leq i \leq r \leq p$ 
$$
E_{\bar{\Sigma}_n} \left[ d_i^r (\bar{\boldsymbol \eta}_n^r) \right] = 0. 
$$
\end{lemma}

\noindent
{\it Proof}: Let $\bar{\Sigma}_{n,r}$ denote the sub matrix of $\bar{\Sigma}_n$ formed by using the first $r$ rows and 
columns. Since $\bar{\Omega}_n = \bar{L}_n^t \bar{L}_n$, it follows that $\bar{\eta}^r_{n,r} 
\bar{\boldsymbol \eta}_n^r$ is the $r^{th}$ row of $\left( \bar{\Sigma}_{n,r} \right)^{-1}$. It follows that for every 
$1 \leq i < r \leq p$, 
$$
E_{\bar{\Sigma}_n} \left[ d_i^r (\bar{\eta}_n^r) \right] = 2 \sum_{j=1}^r \bar{\eta}_{n,j}^r \bar{\Sigma}_{n,ij} = 
\frac{2}{\bar{\eta}^r_r} \sum_{j=1}^r \left( \bar{\Sigma}_{n,r} \right)^{-1}_{rj} \left( \bar{\Sigma}_{n,r} \right)_{ij} = 
0, 
$$

\noindent
and 
$$
E_{\bar{\Sigma}_n} \left[ d_r^r (\bar{\eta}_n^r) \right] = \frac{2}{\bar{\eta}^r_r} \sum_{j=1}^r \left( \bar{\Sigma}_{n,r} 
\right)^{-1}_{rj} \left( \bar{\Sigma}_{n,r} \right)_{rj} - \frac{2}{\bar{\eta}_r^r} = 0. 
$$

\noindent
\hfill$\Box$ 

%
%
%

\begin{lemma} \label{lem:consistency:5}
For any $\gamma > 0$, there exists a constant $C_{1, \gamma} > 0$ such that with probability at least $1 - 
O(n^{-\gamma})$, 
$$
\max_{1 \leq i \leq r \leq p} |d_i^r (\bar{\boldsymbol \eta}_n^r)|  \leq C_{1, \eta} \sqrt{\frac{\log n}{n}}. 
$$
\end{lemma}

\noindent
{\it Proof}: Note that 
$$
d_i^r (\bar{\boldsymbol \eta}_n^r) = \frac{2}{n} \sum_{\ell=1}^n Y^\ell_i (\sum_{j=1}^r \bar{\eta}_{n,j}^r Y^\ell_j) 
- \frac{2}{\bar{\eta}_r^r} 1_{\{i = r\}} 
$$

\noindent
is the difference between the sample covariance and population covariance of $Y_i$ and $\sum_{j=1}^r 
\bar{\eta}_{n,j}^r Y_j$ Since $\bar{\eta}^r_{n,r} \bar{\boldsymbol \eta}_n^r$ is the $r^{th}$ row of $\left( 
\bar{\Sigma}_{n,r} \right)^{-1}$, it follows by Assumption (A1) that the variance of $\sum_{j=1}^r 
\bar{\eta}_{n,j}^r Y_j$, given by $(\bar{\boldsymbol \eta}^r)^t \Sigma_{n,r} \bar{\boldsymbol \eta}$ is 
uniformly bounded over $n$ and $r$. The proof now follows along the same lines as in the proof of 
Lemma \ref{lem:consistency:1}. \hfill$\Box$ 

\bigskip

\noindent
With the above lemmas in hand, we now move towards the main proof. Fix $r$ between $2$ and $p$ 
arbitrarily. Recall that $\mathcal{A}_n^r$ (henceforth referred to as $\mathcal{A}^r$) is the set of indices 
corresponding to the non-zero entries of $\bar{\boldsymbol \eta}_n^r$. We start by establishing 
properties for the following restricted minimization problem: 
\begin{equation} \label{eq20}
\mbox{Minimize } Q_{CSCS,r}({\boldsymbol \eta}^r) \mbox{ w.r.t. } {\boldsymbol \eta}^r \mbox{ such that } 
\eta^r_j = 0 \mbox{ for every } j \notin \mathcal{A}^r. 
\end{equation}

\begin{lemma} \label{lem:consistency:6.1}
There exists $C > 0$ such that for any $\gamma > 0$, a global minima of the restricted minimization problem 
in (\ref{eq20}) exists within the disc $\{{\boldsymbol \eta}^r: \|{\boldsymbol \eta}^r - \bar{\boldsymbol \eta}^r\| < 
C \sqrt{|\mathcal{A}^r|} \lambda_n\}$ with probability at least $1 - O(n^{-\gamma})$ for sufficiently large $n$. 
\end{lemma}

\noindent
{\it Proof}: Let $\alpha_n = \sqrt{|\mathcal{A}^r|} \lambda_n$. Then for any constant $C > 0$ and any 
${\bf u} \in \mathbb{R}^r$ satisfying $u_j = 0$ for every $j \notin \mathcal{A}^r$ and $\|{\bf u}\| = C$, we 
get by the triangle inequality that 
\begin{equation} \label{eq21}
\sum_{j=1}^{r-1} |\bar{\eta}^r_j| - \sum_{j=1}^{r-1} |\bar{\eta}^r_j + \alpha_n u_j| \leq \alpha_n \sum_{j=1}^{r-1} 
|u_j| \leq C \alpha_n \sqrt{|\mathcal{A}^r|}. 
\end{equation}

\noindent
Let 
$$
\tilde{Q}_{CSCS,r} ({\boldsymbol \eta}^r) :=  ({\boldsymbol \eta}^r)^T S_r {\boldsymbol \eta}^r - 2 
\log \eta^r_r. 
$$

\noindent
By (\ref{eq21}) and a second order Taylor series expansion around $\bar{\boldsymbol \eta}^r$, we get 
\begin{eqnarray}
& & Q_{CSCS,r} (\bar{\boldsymbol \eta}^r + \alpha_n {\bf u}) - Q_{CSCS,r} (\bar{\boldsymbol \eta}^r) 
\nonumber\\
&=& \tilde{Q}_{CSCS,r} (\bar{\boldsymbol \eta}^r + \alpha_n {\bf u}) - \tilde{Q}_{CSCS,r} 
(\bar{\boldsymbol \eta}^r)  - \lambda_n \left( \sum_{j=1}^{r-1} |\bar{\eta}^r_j| - \sum_{j=1}^{r-1} |\bar{\eta}^r_j 
+ \alpha_n u_j| \right) \nonumber\\
&\geq& \alpha_n \sum_{j \in \mathcal{A}^r} u_j d_j^r (\bar{\boldsymbol \eta}^r) + \alpha_n^2 
\sum_{j \in \mathcal{A}^r} \sum_{k \in \mathcal{A}^r} u_j u_k S_{jk} + \frac{u_r^2}{2 (\eta^r_*)^2} - 
C \alpha_n \sqrt{|\mathcal{A}^r|} \lambda_n \nonumber\\
&\geq& \alpha_n \sum_{j \in \mathcal{A}^r} u_j d_j^r (\bar{\boldsymbol \eta}^r) + \alpha_n^2 
\sum_{j \in \mathcal{A}^r} \sum_{k \in \mathcal{A}^r} u_j u_k (S_{jk} - \bar{\Sigma}_{n,jk}) + \sum_{j \in 
\mathcal{A}^r} \sum_{k \in \mathcal{A}^r} u_j u_k \bar{\Sigma}_{n, jk} - C \alpha_n^2 \label{eq22}
\end{eqnarray}

\noindent
where $\eta^r_* \in [\bar{\eta}_r^r, \bar{\eta}_r^r + \alpha_n u_r]$. Note that $\lambda_n \sqrt{\frac{n}{\log 
n}} \rightarrow \infty$, and $q_n \sqrt{\frac{\log n}{n}} \rightarrow 0$ as $n \rightarrow \infty$, by 
Assumption (A1). It follows by Cauchy-Schwarz inequality, Lemma \ref{lem:consistency:1} and 
Lemma \ref{lem:consistency:5} that for any $\gamma > 0$, there exist constants $C_\gamma$ and 
$C_{1, \gamma} > 0$ such that with probability at least $1 - O(n^{-\gamma})$, 
\begin{equation} \label{eq23}
\alpha_n \sum_{j \in \mathcal{A}^r} u_j d_j^r (\bar{\boldsymbol \eta}^r) \leq C C_{1, \gamma} 
\sqrt{\frac{|\mathcal{A}^r| \log n}{n}} \alpha_n = o(\alpha_n^2), 
\end{equation}

\noindent
and 
\begin{equation} \label{eq24}
\frac{\alpha_n^2}{2} |\sum_{j \in \mathcal{A}^r} \sum_{k \in \mathcal{A}^r} u_j u_k (S_{jk} - \bar{\Sigma}_{n,jk})| 
\leq C_\gamma C^2 q_n \sqrt{\frac{\log n}{n}} = o(\alpha_n^2). 
\end{equation}

\noindent
Also, by Assumption A1, it follows that 
\begin{equation} \label{eq25}
\sum_{j \in \mathcal{A}^r} \sum_{k \in \mathcal{A}^r} u_j u_k \bar{\Sigma}_{n, jk} \geq \frac{C^2 
\alpha_n^2}{2 \lambda_{max}}. 
\end{equation}

\noindent
Combining (\ref{eq22}), (\ref{eq23}), (\ref{eq24}) and (\ref{eq25}), we get that 
$$
Q_{CSCS,r} (\bar{\boldsymbol \eta}^r + \alpha_n {\bf u}) - Q_{CSCS,r} (\bar{\boldsymbol \eta}^r)  > 
\frac{C^2 \alpha_n^2}{2 \lambda_{max}} - 2C \alpha_n^2 
$$

\noindent
with probability at least $1 - O(n^{-\gamma})$ for large enough $n$. Choosing $C = 4 \lambda_{max} + 1$, 
we obtain that 
$$
\inf_{{\bf u}: {\bf u}_j = 0 for j \notin \mathcal{A}^r, \|{\bf u}\| = C} Q_{CSCS,r} (\bar{\boldsymbol \eta}^r + 
\alpha_n {\bf u}) > Q_{CSCS,r} (\bar{\boldsymbol \eta}^r),  
$$

\noindent
with probability at least $1 - O(n^{-\gamma})$ for large enough$n$. Hence for every $\eta > 0$, a local 
minima (in fact global minima due to convexity) of the restricted minimization problem in (\ref{eq20}) exists 
within the disc $\{{\boldsymbol \eta}^r: \|{\boldsymbol \eta}^r - \bar{\boldsymbol \eta}^r\| < C 
\sqrt{|\mathcal{A}^r|} \lambda_n\}$ with probability at least $1 - O(n^{-\eta})$ for sufficiently large $n$. 
\hfill$\Box$ 

\begin{lemma} \label{lem:consistency:6}
There exists a constant $C_1 > 0$, such that for any $\gamma > 0$ the following holds with probability 
$1 - O(n^{-\gamma})$: for any ${\boldsymbol \eta}^r$ in the set 
$$
S = \{{\boldsymbol \eta}^r: \|{\boldsymbol \eta}^r - \bar{\boldsymbol \eta}^r\| \geq C_1 \sqrt{|\mathcal{A}^r|} 
\lambda_n, \; \eta_j^r = 0 \; \forall j \notin \mathcal{A}^r\}, 
$$ 

\noindent
we have $\left\| {\bf d}_{\mathcal{A}^r}^r (\bar{\boldsymbol \eta}^r) \right\| > \sqrt{|\mathcal{A}^r|} \lambda_n$, 
where ${\bf d}_{\mathcal{A}^r}^r (\bar{\boldsymbol \eta}^r) := \left( d_j^r (\bar{\boldsymbol \eta}^r) \right)_{j \in 
\mathcal{A}^r}$. 
\end{lemma}

\noindent
{\it Proof}: Recall that $\alpha_n = \sqrt{|\mathcal{A}^r|} \lambda_n$. Choose ${\boldsymbol \eta} \in S$ 
arbitrarily. Let ${\bf u} = {\boldsymbol \eta}^r - \bar{\boldsymbol \eta}^r/\alpha_n$. It follows that $u_j = 0$ 
for every $j \notin \mathcal{A}^r$, and $\|{\bf u}\| \geq C_1$. Let $\Delta_r$ denote the $|\mathcal{A}^r| 
\times |\mathcal{A}^r|$ matrix with the diagonal entry corresponding to the $r^{th}$ variable equal to $1$, 
and all other entries equal to zero. By a first order Taylor series expansion ${\bf d}_{\mathcal{A}^r}^r$, it 
follows that 
\begin{eqnarray}
{\bf d}_{\mathcal{A}^r}^r ({\boldsymbol \eta}^r) 
&=& {\bf d}_{\mathcal{A}^r}^r (\bar{\boldsymbol \eta}^r) + 2 \alpha_n \left( S_{\mathcal{A}^r \mathcal{A}^r} + 
\frac{1}{({\boldsymbol \eta}^r_*)^2} \Delta_r \right) {\bf u}_{\mathcal{A}^r} \nonumber\\
&=& {\bf d}_{\mathcal{A}^r}^r (\bar{\boldsymbol \eta}^r) + 2 \alpha_n \left( \Sigma_{\mathcal{A}^r 
\mathcal{A}^r} + \frac{1}{(\eta_*^r)^2} \Delta_r \right) {\bf u}_{\mathcal{A}^r} + 2 \alpha_n 
\left( S_{\mathcal{A}^r \mathcal{A}^r} - \Sigma_{n, \mathcal{A}^r \mathcal{A}^r} \right) 
{\bf u}_{\mathcal{A}^r}, \nonumber\\
& & \label{eq26}
\end{eqnarray}

\noindent
where $\eta_r^*$ lies between $\bar{\eta}_r^r$ and $\bar{\eta}_r^r + \alpha_n u_r$. By 
Lemma \ref{lem:consistency:1} and Lemma \ref{lem:consistency:5}  it follows that for any $\gamma > 0$, 
there exist constants $C_{2, \gamma}$ and $C_{3, \gamma}$ such that 
\begin{eqnarray*}
& & \| {\bf d}_{\mathcal{A}^r}^r ({\boldsymbol \eta}^r) \|\\
&\geq& 2 \alpha_n \left\| \left( \Sigma_{\mathcal{A}^r \mathcal{A}^r} + \frac{1}{({\boldsymbol \eta}^r_*)^2} 
\Delta_r \right) {\bf u}_{\mathcal{A}^r} \right\| - C_{2, \gamma} \sqrt{\frac{q_n \log n}{n}} - C_{3, \gamma} 
\|{\bf u}\| \frac{\alpha_n |\mathcal{A}^r| \sqrt{\log n}}{\sqrt{n}}\\
&\geq& \frac{\alpha_n}{\lambda_{max}} \|{\bf u}\| = \sqrt{|\mathcal{A}^r|} \lambda_n 
\frac{C_1}{\lambda_{max}} 
\end{eqnarray*}

\noindent
with probability at least $1 - O(n^{-\gamma})$, for large enough $n$. The last inequality follows from 
Assumption (A1), the fact that $|\mathcal{A}^r| \leq q_n$ and Assumption (A5). Choosing $C_1 = 
\lambda_{max} + 1$ leads to the required result. \hfill$\Box$ 

\medskip

\noindent
The next lemma establishes estimation and model selection (sign) consistency for the restricted 
minimization problem in (\ref{eq20}). 
\begin{lemma} \label{lem:consistency:7}
There exists $C_2 > 0$ such that for any $\gamma > 0$, the following holds with probability $1 - 
O(n^{-\gamma})$, for large enough $n$: (i) there exists a solution to the minimization problem in 
(\ref{eq20}), (ii) (estimation consistency) any global minimum of the restricted minimization problem in 
(\ref{eq20}) lies within the disc $\{{\boldsymbol \eta}^r: \|{\boldsymbol \eta}^r - \bar{\boldsymbol \eta}^r\| < 
C_2 \sqrt{|\mathcal{A}^r|} \lambda_n\}$, and (iii) (sign consistency) for any solution 
$\hat{\boldsymbol \eta}^r$ of the minimization problem in (\ref{eq20}),  $\mbox{sign}(\hat{\eta}_j) = 
\mbox{sign}(\bar{\eta}_j)$ for every $1 \leq j \leq r$.  
\end{lemma}

\noindent
{\it Proof}: The existence  of a solution follows from Lemma \ref{lem:consistency:6}. By the KKT conditions 
for the restricted minimization problem in (\ref{eq20}) (along the lines of Lemma \ref{lem:consistency:2}), it 
follows that for any solution $\hat{\boldsymbol \eta}^r$ of (\ref{eq20}), $|d_j^r (\hat{\boldsymbol \eta}^r)| \leq 
\lambda_n$ for every $j \in \mathcal{A}^r$. It follows that $\left\| {\bf d}_{\mathcal{A}^r}^r 
(\hat{\boldsymbol \eta}^r) \right\| \leq \sqrt{\mathcal{A}^r} \lambda_n$. The estimation consistency now 
follows from Lemma \ref{lem:consistency:7}. Note that by Assumption (A4), $\bar{\eta}_j^r \geq s_n > 
2 C_2 \sqrt{|\mathcal{A}^r|} \lambda_n$ for every $j \in \mathcal{A}^r$, for sufficiently large $n$. The sign 
consistency now follows by combining this fact with $\|{\boldsymbol \eta}^r - \bar{\boldsymbol \eta}^r\| < 
C_2 \sqrt{|\mathcal{A}^r|} \lambda_n$. \hfill$\Box$ 

\medskip 

\noindent
The next lemma will be instrumental in showing that the solution set of the restricted minimization problem 
in (\ref{eq20}) is the same as the solution set of the unrestricted minimization problem for $Q_{CSCS,r}$ 
with high probability. 
\begin{lemma} \label{lem:consistency:8}
For any $\gamma > 0$, any solution $\hat{\boldsymbol \eta}^r$ of (\ref{eq20}) satisfies 
$$
\max_{j \notin \mathcal{A}^r} \left| d_j^r (\hat{\boldsymbol \eta}^r) \right| < \lambda_n 
$$

\noindent
with probability at least $1 - O(n^{-\gamma})$, for large enough $n$. 
\end{lemma}

\noindent
{\it Proof}: Let $\gamma > 0$ be given, and let $\hat{\boldsymbol \eta}^r$ be a solution of (\ref{eq20}). If 
$C_n := \{\mbox{sign}(\hat{\boldsymbol \eta}^r) = \mbox{sign}(\bar{\boldsymbol \eta}^r)\}$, then $P(C_n) 
\geq 1 - O(n^{-\gamma-\kappa})$ for large enough $n$ (by Lemma \ref{lem:consistency:7}). Now, on 
$C_n$, it follows by the a first order expansion of ${\bf d}_{\mathcal{A}^r}^r$ around 
$\bar{\boldsymbol \eta}^r$, and the KKT conditions for (\ref{eq20}) that 
\begin{eqnarray}
- \lambda_n \mbox{sign} \left( \bar{\boldsymbol \eta}_{\mathcal{A}^r}^r \right) 
&=& {\bf d}_{\mathcal{A}^r}^r (\hat{\boldsymbol \eta}^r) \nonumber\\
&=& {\bf d}_{\mathcal{A}^r}^r (\bar{\boldsymbol \eta}^r) + 2 S_{\mathcal{A}^r \mathcal{A}^r} \hat{\bf u}_n 
+ \frac{2}{(\eta_r^*)^2} \Delta_r \hat{\bf u}_n \nonumber\\
&=& H_n \hat{\bf u}_n + {\bf d}_{\mathcal{A}^r}^r (\bar{\boldsymbol \eta}^r) + 2 \left( S_{\mathcal{A}^r 
\mathcal{A}^r} - \bar{\Sigma}_{n, \mathcal{A}^r \mathcal{A}^r} \right) \hat{\bf u}_n + \nonumber\\
& & \left( \frac{2}{(\eta_r^*)^2} - \frac{2}{(\bar{\eta}_r^r)^2} \right) \Delta_r \hat{\bf u}_n, \label{eq27} 
\end{eqnarray}

\noindent
where $\hat{\bf u}_n = \hat{\boldsymbol \eta}^r - \bar{\boldsymbol \eta}^r$, $\eta_r^*$ lies between 
$\bar{\eta}_r^r$ and $\hat{\eta}_r^r$, and $H_n = 2 \bar{\Sigma}_{n, \mathcal{A}^r \mathcal{A}^r} + 
\frac{2}{(\bar{\eta}_r^r)^2} \Delta_r$. Hence, 
\begin{eqnarray} \label{eq28}
\hat{\bf u}_n 
&=& - \lambda_n H_n^{-1} \mbox{sign} \left( \bar{\boldsymbol \eta}_{\mathcal{A}^r}^r \right) - H_n^{-1} 
{\bf d}_{\mathcal{A}^r}^r (\bar{\boldsymbol \eta}^r) - 2 H_n^{-1} \left( S_{\mathcal{A}^r \mathcal{A}^r} - 
\bar{\Sigma}_{n, \mathcal{A}^r \mathcal{A}^r} \right) \hat{\bf u}_n \nonumber\\
& & - 2 H_n^{-1} \left( \frac{1}{(\eta_r^*)^2} - \frac{1}{(\bar{\eta}_r^r)^2} \right) \Delta_r \hat{\bf u}_n.  
\end{eqnarray}

\noindent
Now, let us fix $j \notin \mathcal{A}^r$. By a first order Taylor series expansion of $d_j^r$, it follows 
that 
$$
d_j^r (\hat{\boldsymbol \eta}^r) = d_j^r (\bar{\boldsymbol \eta}^r) + 2 S_{i,\mathcal{A}^r}^t \hat{\bf u}_n. 
$$

\noindent
Using (\ref{eq28}), we get that 
\begin{eqnarray} \label{eq29}
d_j^r (\hat{\boldsymbol \eta}^r) 
&=& d_j^r (\bar{\boldsymbol \eta}^r) + 2 (S_{j,\mathcal{A}^r} - \bar{\Sigma}_{n,j,\mathcal{A}^r})^t 
\hat{\bf u}_n + 2 \bar{\Sigma}_{n,j,\mathcal{A}^r}^t \hat{\bf u}_n \nonumber\\
&=&  -2 \lambda_n \bar{\Sigma}_{n,j,\mathcal{A}^r}^t H_n^{-1} \mbox{sign} \left( 
\bar{\boldsymbol \eta}_{\mathcal{A}^r}^r \right) + d_j^r (\bar{\boldsymbol \eta}^r) - 2 
\bar{\Sigma}_{n,j,\mathcal{A}^r}^t H_n^{-1} {\bf d}_{\mathcal{A}^r}^r (\bar{\boldsymbol \eta}^r) + 
\nonumber\\
& & - 4 \bar{\Sigma}_{n,j,\mathcal{A}^r}^t H_n^{-1} \left( S_{\mathcal{A}^r \mathcal{A}^r} - 
\bar{\Sigma}_{n, \mathcal{A}^r \mathcal{A}^r} \right) \hat{\bf u}_n - 4 \Sigma_{n,j,\mathcal{A}^r}^t 
H_n^{-1} \left( \frac{1}{(\eta_r^*)^2} - \frac{1}{(\bar{\eta}_r^r)^2} \right) \Delta_r \hat{\bf u}_n + 
\nonumber\\
& & 2 (S_{i,\mathcal{A}^r} - \bar{\Sigma}_{n,i,\mathcal{A}^r})^t \hat{\bf u}_n. 
\end{eqnarray}

\noindent
We now individually analyze all the terms in (\ref{eq29}). It follows by the ``incoherence" 
Assumption (A3) that the first term satisfies 
\begin{equation} \label{eq30}
\left| -2 \lambda_n \bar{\Sigma}_{n,j,\mathcal{A}^r}^t H_n^{-1} \mbox{sign} \left( 
\bar{\boldsymbol \eta}_{\mathcal{A}^r}^r \right) \right| \leq \delta \lambda_n < \lambda_n. 
\end{equation}

\noindent
It follows by Lemma \ref{lem:consistency:5} and Assumption (A5) that the second term $d_j^r 
(\bar{\boldsymbol \eta}^r)$ is $o(\lambda_n)$ with probability $1 - O(n^{-\gamma-\kappa})$ for 
large enough $n$. Also, by Assumption (A1) and the definition of $H_n$, we get that 
\begin{equation} \label{eq31}
\left\| 2 \bar{\Sigma}_{n,j,\mathcal{A}^r}^t H_n^{-1} \right\| \leq \left\| \bar{\Sigma}_{n,j,\mathcal{A}^r} 
\right\| \| 2 H_n^{-1} \| \leq \frac{1}{\lambda_{min}} \left\| \Sigma_{n, \mathcal{A}^r \mathcal{A}^r}^{-1} 
\right\| \leq \frac{\lambda_{max}}{\lambda_{min}}. 
\end{equation}

\noindent
It follows by Lemma \ref{lem:consistency:5} and Assumption (A5) that the third term in (\ref{eq29}) 
satisfies 
\begin{equation} \label{eq32}
\left| 2 \bar{\Sigma}_{n,j,\mathcal{A}^r}^t H_n^{-1} {\bf d}_{\mathcal{A}^r}^r (\bar{\boldsymbol 
\eta}^r) \right| \leq \frac{\lambda_{max}}{\lambda_{min}} \sqrt{q_n} \max_{j \in \mathcal{A}^r} 
|d_j^r (\bar{\boldsymbol \eta}^r)| = o(\lambda_n). 
\end{equation}

\noindent
Let ${\bf b} = 2 H_n^{-1} \Sigma_{n,j,\mathcal{A}^r}$. Note that by (\ref{eq31}), the norm of 
$\{\bf b\}$ is uniformly bounded in $n$ and $r$. Note that the $j^{th}$ element of the vector $\left( 
S_{\mathcal{A}^r \mathcal{A}^r} - \bar{\Sigma}_{n, \mathcal{A}^r \mathcal{A}^r} \right) {\bf b}$ 
is the difference between the sample and the population covariance of $Y_j$ and $\sum_{k \in 
\mathcal{A}^r} b_k Y_k$. Using the same line of arguments as in the proof of 
Lemma \ref{lem:consistency:5}, it follows that there exists a constant $C_{4, \gamma} > 0$ such 
that 
\begin{equation} \label{eq33}
\max_{j \in \mathcal{A}^r} \left| \left( \left( S_{\mathcal{A}^r \mathcal{A}^r} - \bar{\Sigma}_{n, 
\mathcal{A}^r \mathcal{A}^r} \right) {\bf b} \right)_j \right| \leq C_{4, \gamma} 
\sqrt{\frac{\log n}{n}}, 
\end{equation}

\noindent
with probability $1 - O(n^{-\gamma-\kappa})$, for large enough $n$. By (\ref{eq31}), (\ref{eq33}), 
the estimation consistency part of Lemma \ref{lem:consistency:7} and Assumption (A5) that the 
fourth term in (\ref{eq29}) satisfies 
\begin{eqnarray} \label{eq34}
\left| 4 \bar{\Sigma}_{n,j,\mathcal{A}^r}^t H_n^{-1} \left( S_{\mathcal{A}^r \mathcal{A}^r} - 
\bar{\Sigma}_{n, \mathcal{A}^r \mathcal{A}^r} \right) \hat{\bf u}_n \right| 
&\leq& 2 \left\| \left( S_{\mathcal{A}^r \mathcal{A}^r} - \bar{\Sigma}_{n, \mathcal{A}^r \mathcal{A}^r} 
\right) {\bf b} \right\| \|\hat{\bf u}_n\| \nonumber\\ 
&=& O \left( \sqrt{\frac{|\mathcal{A}^r \log n}{n}} \sqrt{|\mathcal{A}^r|} \lambda_n \right) = 
o(\lambda_n), 
\end{eqnarray}

\noindent
with probability $1 - O(n^{-\gamma-\kappa})$, for large enough $n$. Since $(\bar{\eta}_r^r)^2$ is the 
$r^{th}$ diagonal entry of $\bar{\Sigma}_{n,r}^{-1}$, it follows by Assumption (A1) that $\bar{\eta}_r^r$ 
is uniformly bounded above and below in $n$ and $r$. Since $\eta_r^*$ lies between $\bar{\eta}_r^r$ 
and $\hat{\eta}_r^r$, it follows by the estimation consistency part of Lemma \ref{lem:consistency:7} 
that $\eta_r^*$ is bounded above and below uniformly with probability at least $1 - 
O(n^{-\gamma-\kappa})$, for large enough $n$. By (\ref{eq31}), the definition of $\Delta_r$, 
Lemma \ref{lem:consistency:7}, and Assumption (A1), the fifth term in (\ref{eq29}) satisfies 
\begin{eqnarray} \label{eq35}
\left| 4 \Sigma_{n,j,\mathcal{A}^r}^t H_n^{-1} \left( \frac{1}{(\eta_r^*)^2} - \frac{1}{(\bar{\eta}_r^r)^2} \right) 
\Delta_r \hat{\bf u}_n \right| 
&\leq& \frac{2 \lambda_{max} |\bar{\eta}_r^r + \eta_r^*| |}{\lambda_{min} (\bar{\eta}_r^r)^2 (\eta_r^*)^2} 
|\bar{\eta}_r^r - \hat{\eta}_r^r| |\hat{u}_{n,r}| \nonumber\\
&=& O \left( |\mathcal{A}^r| \lambda_n^2 \right) = o(\lambda_n). 
\end{eqnarray}

\noindent
with probability $1 - O(n^{-\gamma-\kappa})$, for large enough $n$. Also, by 
Lemma \ref{lem:consistency:1}, the consistency part of Lemma \ref{lem:consistency:7}, and 
Assumption (A1), the sixth term in (\ref{eq29}) satisfies 
\begin{equation} \label{eq36}
\left| 2 (S_{i,\mathcal{A}^r} - \bar{\Sigma}_{n,i,\mathcal{A}^r})^t \hat{\bf u}_n \right| \leq 2 \left\| 
S_{i,\mathcal{A}^r} - \bar{\Sigma}_{n,i,\mathcal{A}^r} \right\| \|\hat{\bf u}_n\| = O \left( 
\sqrt{\frac{|\mathcal{A}^r| \log n}{n}} \sqrt{|\mathcal{A}^r|} \lambda_n \right) = o(\lambda_n). 
\end{equation}

\noindent
It follows by (\ref{eq29}), (\ref{eq30}), (\ref{eq32}), (\ref{eq34})-(\ref{eq36}) that for any $j \notin 
\mathcal{A}^r$, 
$$
\left| d_j^r (\hat{\boldsymbol \eta}^r) \right| < \lambda_n 
$$

\noindent
with probability at least $1 - O(n^{-\gamma-\kappa})$, for large enough $n$. The result now 
follows by the union bound, and from the fact that $p = O(n^\kappa)$. \hfill$\Box$ 

\medskip

\noindent
Let $\gamma > 0$ and $1 \leq r \leq p$ be chosen arbitrarily. Let $C_{r,n}$ denote the event on which 
Lemma \ref{lem:consistency:7} and Lemma \ref{lem:consistency:8} hold. It follows that 
$P(C_{r,n}) \geq 1 - O(n^{-\gamma-\kappa})$, for large enough $n$. Now, on $C_{r,n}$, any solution 
of the restricted problem (\ref{eq20}) is also a global minimizer of $Q_{CSCS,r}$ (by 
Lemma \ref{lem:consistency:2}). Hence, there is at least one global minimizer of $Q_{CSCS,r}$ 
for which the components corresponding to $(\mathcal{A}^r)^c$ are zero. It again follows by 
Lemma \ref{lem:consistency:2} that these components are zero for all global minimizers 
of $Q_{CSCS,r}$. Hence, the solution set of the restricted minimization problem in 
(\ref{eq20}) is the same as the solution set for the unrestricted problem (i.e., the set of 
global minimizers of $Q_{CSCS,r}$). Hence, on $C_{r,n}$, the assertions of 
Lemma \ref{lem:consistency:7} hold for the solutions of the unrestricted minimization problem 
for $Q_{CSCS,r}$. 

Recall that $Q_{CSCS} (L) = \sum_{r=1}^p Q_{CSCS, r} ({\boldsymbol \eta}^r)$, and that 
$\{{\boldsymbol \eta}^r\}_{r=1}^p$ form a disjoint partition of $L$. Note that by the union bound and 
the fact that $p = O(n^\kappa)$, $P(\cap_{r=1}^n C_{r,n}) \geq 1 - O(n^{-\gamma})$, for large 
enough $n$. Also, by the triangle inequality $\| L - \widetilde{L} \| \leq \sum_{r=1}^p \| 
{\boldsymbol \eta}^r - \widetilde{\boldsymbol \eta}^r \|$ for any $L, \widetilde{L} \in \mathcal{L}_p$. It 
follows that the assertions in Theorem \ref{thm:consistency} hold on $\cap_{r=1}^p C_{r,n}$. 
\hfill$\Box$ 

\section{Call center data: forecasting details} \label{ccforecast}

\noindent
Suppose $y_i = (y_{i,1},...,y_{i,102})'$, and $y_i = (y_i^{(1)'}, y_i^{(2)'} ) '$, where 
$y_i^{(1)}$ and $y_i^{(2)}$ are $51$ dimensional vectors that measure the arrival 
patterns in the early and later times of day $i$. The corresponding partitions for the 
mean and covariance matrix are denoted by $\mu' = (\mu_1', \mu_2')$ and
\begin{align*}
\Sigma = \begin{pmatrix} \Sigma_{11} & \Sigma_{12} \\
\Sigma_{21} & \Sigma_{22}
\end{pmatrix} 
\end{align*}

Assuming multivariate normality, the best mean squared error forecast
of $y_i^{(2)}$ using $y_i^{(1)}$ is
\begin{equation} \label{eq:forecast}
\mathbb{E}(y_i^{(2)}|y_i^{(1)}) = \mu_2 + \Sigma_{21} \Sigma_{11}^{-1}
(y_i^{(1)} - \mu_1)
\end{equation}

To compare the forecast performance using the four different covariance matrix 
estimates, we split the 239 days into training and test datasets. The data from the
first $T$ days ($T = 205, 150, 100, 75$), form the training dataset that is used to 
estimate the mean and covariance structure. The estimates are then applied for 
forecasting using (\ref{eq:forecast}) for the $239 - T$ days in the test set. Note 
that we use the 51 square-root-transformed arrival counts in the early half of a 
day to forecast the square-root-transformed arrival counts in the later half of the 
day. For each time interval $t = 52, \hdots, 102$, the authors in \cite{HLPL:2006} 
define the forecast error (FE) by: 
$$
\text{FE}_t = \frac{1}{239-T} \sum_{i = T+1}^{239} |{\hat{y}_{it} - y_{it}}| 
$$

\noindent
where $y_{it}$ and $\hat{y}_{it}$ are the observed and forecast values respectively. 

\begin{figure}[H]
\begin{center}
  \includegraphics[scale = 0.8]{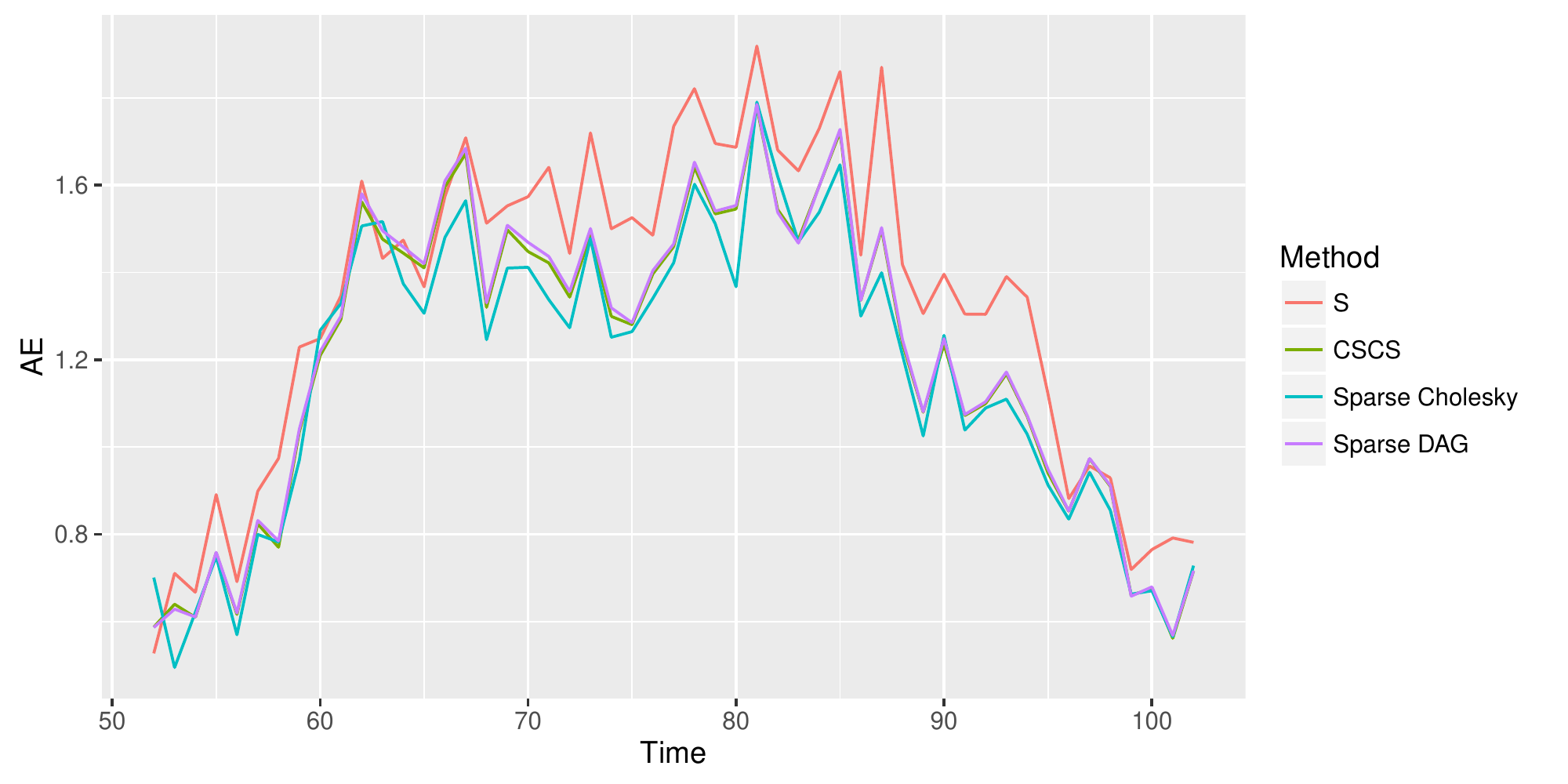}
\end{center}
\caption{Average Absolute Forecast Error with 205 observations in training dataset} 
\label{fig:cccv205}
\end{figure}

\begin{figure}[H]
\begin{center}
  \includegraphics[scale = 0.8]{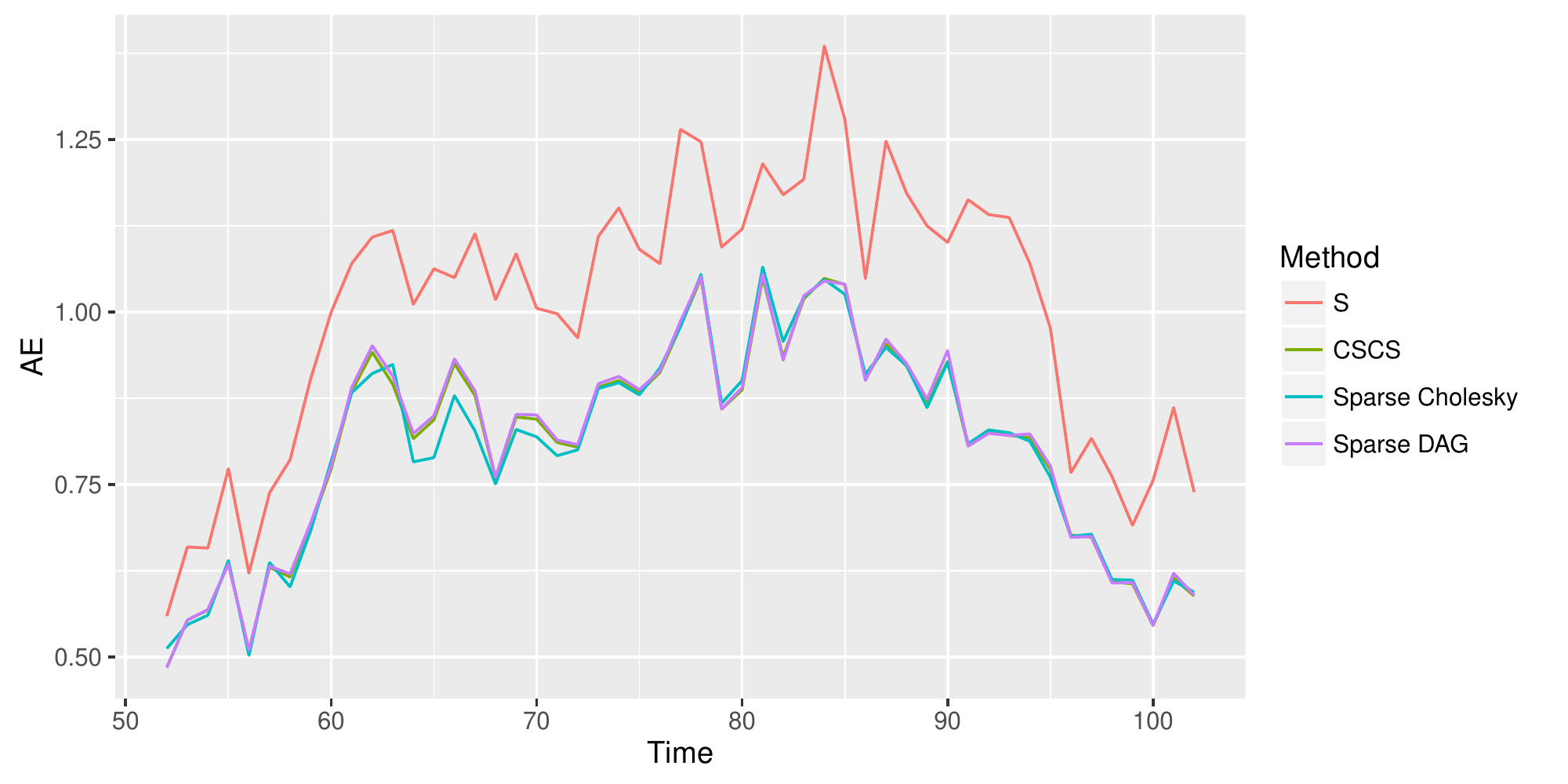}
\end{center}
\caption{Average Absolute Forecast Error with 150 observations in training dataset} 
\label{fig:cccv150}
\end{figure}

\begin{figure}[H]
\begin{center}
  \includegraphics[scale = 0.8]{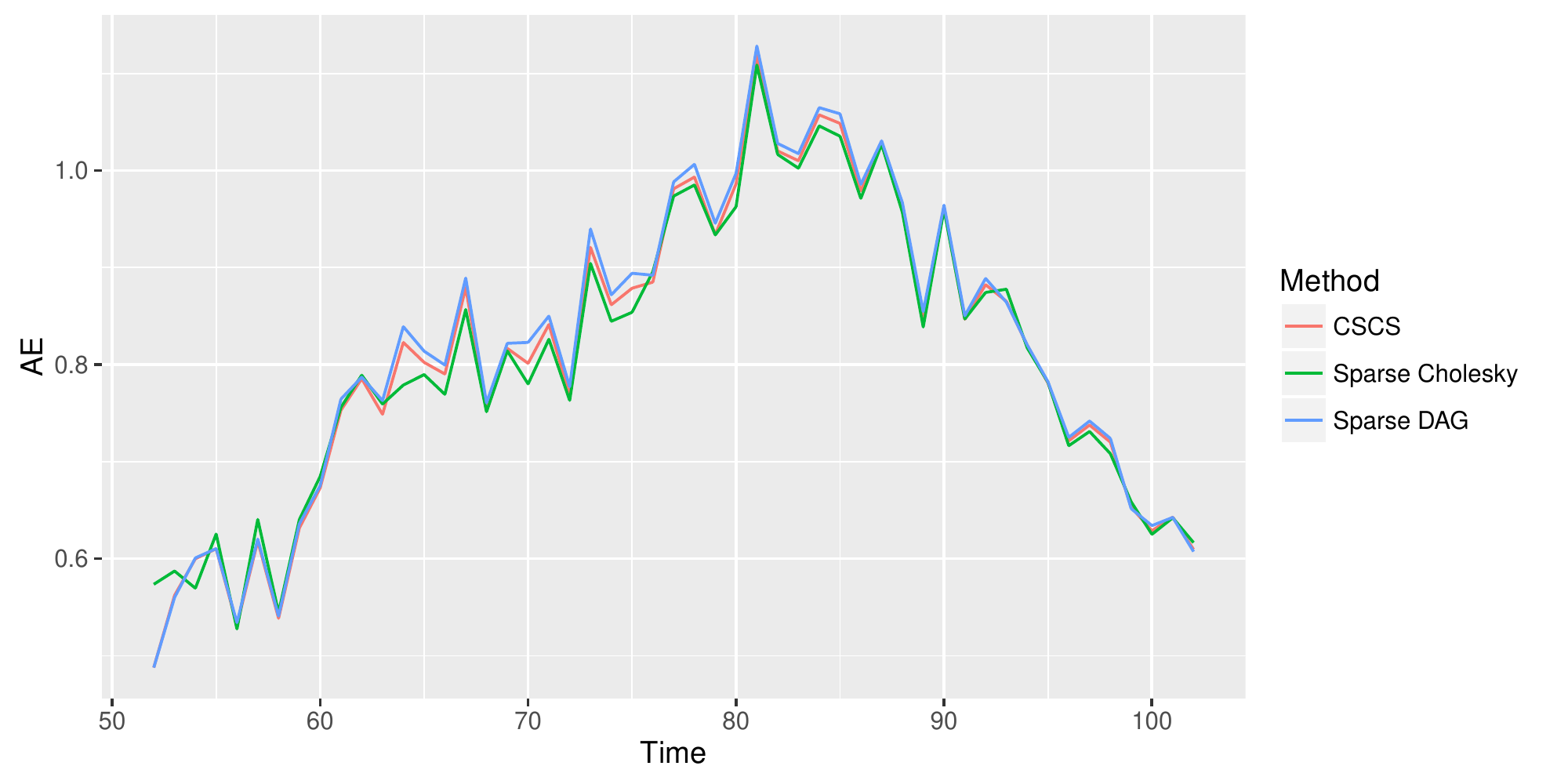}
\end{center}
\caption{Average Absolute Forecast Error with 100 observations in training dataset} 
\label{fig:cccv100}
\end{figure}

\begin{figure}[H]
\begin{center}
  \includegraphics[scale = 0.8]{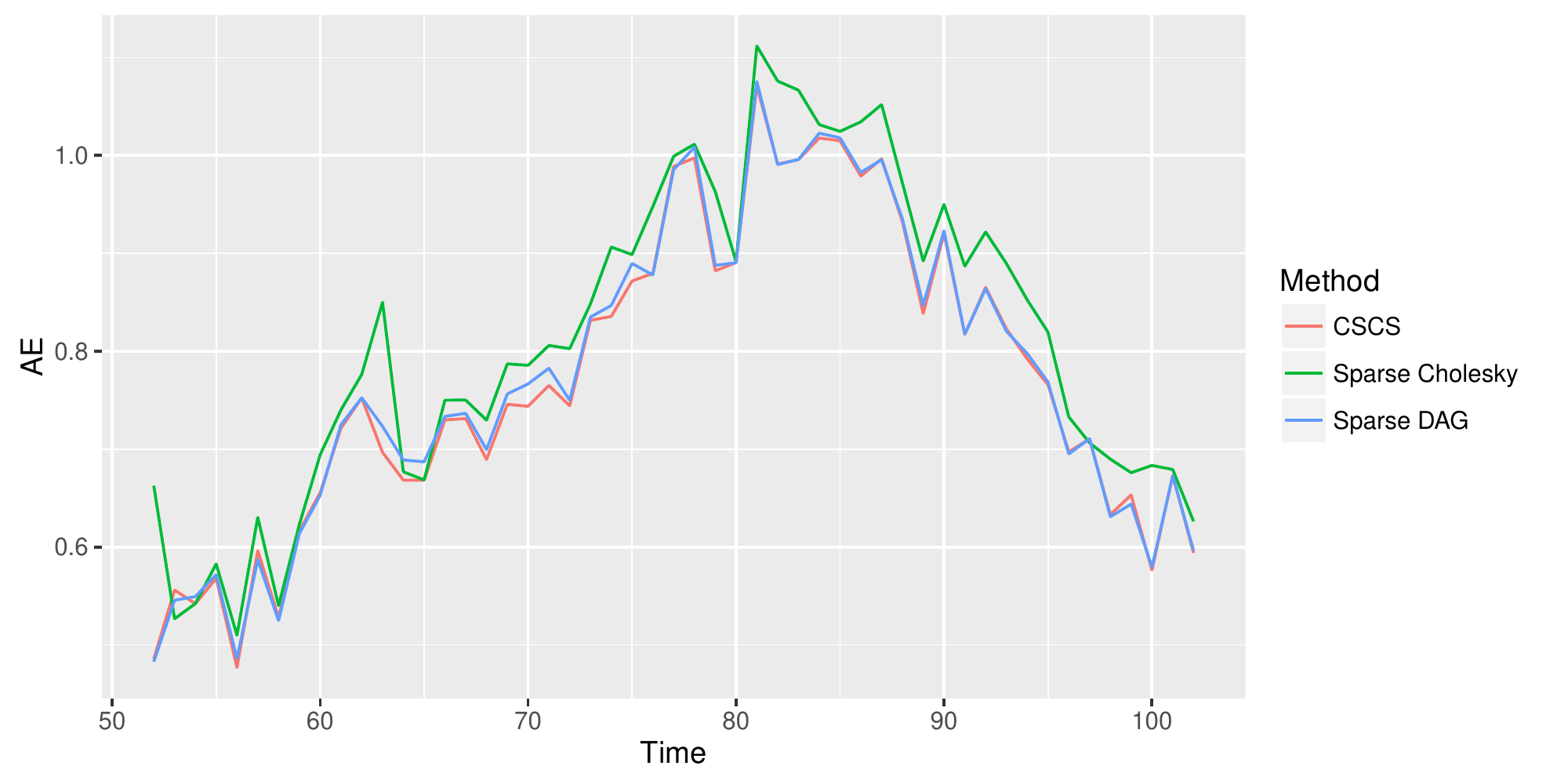}
\end{center}
\caption{Average Absolute Forecast Error with 75 observations in
  training dataset} 
\label{fig:cccv75}
\end{figure}

In Table \ref{tab:aafemin}, we provide the number of time intervals (out of $51$) where 
each of the four methods (CSCS, Sparse Cholesky, Sparse DAG, sample covariance 
matrix) has the minimum forecast error value. Table \ref{tab:aafesum} provides the 
aggregated forecast errors over all the $51$ time intervals for each method. When 
$T = 205$ and the size of the training data is larger than the number of variables, it is 
clear that Sparse Cholesky performs the best, achieving minimum $FE_t$ $38$ times, 
followed by CSCS with $8$ and then Sparse DAG with $3$. This ordering is preserved 
when one looks at the aggregated forecast errors. The sample covariance matrix 
performs the worst and achieves the minimum only twice. A similar pattern is observed 
for $T = 150$. The picture changes quite a bit if we reduce the size of the training 
dataset, especially if it is smaller than the number of variables. In the $n < p$ 
framework ($T = 100, 75$) the performance of CSCS improves drastically as 
compared to Sparse Cholesky and Sparse DAG in terms of the number of times it 
achieves the minimum forecast error as shown in Table \ref{tab:aafemin}. This is also 
supported by the aggregated forecast errors in Table \ref{tab:aafesum}. This highlights 
the fact that CSCS is a useful addition to the collection of sparse Cholesky methods, 
especially when the sample size is smaller than the number of variables. Figures 
\ref{fig:cccv205}, \ref{fig:cccv150}, \ref{fig:cccv100} and \ref{fig:cccv75} below 
provide plots of $\text{FE}_t$ corresponding to the different methods discussed above, 
for varying values of the training data size. 
\begin{table}[H]
\centering
\begin{tabular}{||c|c|c|c|c||}
  \hline
\hline
& \multicolumn{4}{c||}{Training Data Size} \\
\hline
Method & $205$ & $150$ & $100$ & $75$\\ 
  \hline
  CSCS & 8 & 16 & 32 &26 \\ 
  Sparse Cholesky & 38 & 27 & 11 &7 \\ 
  Sparse DAG & 3 & 8 & 8 & 18  \\ 
  S & 2 & 0 & - & -\\
   \hline
\hline
\end{tabular}
\caption{Number of times (out of $51$) each estimation method achieves the 
minimum forecast error for training data size $205, 150, 100, 75$}
\label{tab:aafemin}
\end{table}

\begin{table}[H]
\centering
\begin{tabular}{||c|c|c|c|c||}
  \hline
\hline
& \multicolumn{4}{c||}{Training Data Size} \\
\hline
Method & $205$ & $150$ & $100$ & $75$\\ 
  \hline
  CSCS &60.97049& 41.09635 & 40.51781& 39.21523\\ 
  Sparse Cholesky &59.42691& 40.89093& 40.89374& 41.27573 \\ 
  Sparse DAG &61.7157 & 41.2593& 40.61282& 39.41869\\ 
  S & 67.78564& 51.26088 & - & -\\
   \hline
\hline
\end{tabular}
\caption{Aggregated forecast error for each estimation method for 
training data size $205, 150, 100, 75$}
\label{tab:aafesum}
\end{table}

\section*{References}

\indent

\hangindent=2em
\hangafter=1
[1] J. Friedman, T. Hastie, and R. Tibshirani. Applications of the lasso and grouped 
lasso to the estimation of sparse graphical models. {\it Technical Report, Department 
of Statistics, Stanford University}, 2010. 
\vskip 10pt

\hangindent=2em
\hangafter=1
[2] K. Khare and B. Rajaratnam. Convergence of cyclic coordinatewise l1 
minimization. {\it arxiv}, 2014. 
\vskip 10pt

\hangindent=2em
\hangafter=1
[3] K. Khare, S. Oh, and B. Rajaratnam. A convex pseudo-likelihood framework for 
high dimensional partial correlation estimation with convergence guarantees. {\it 
Journal of the Royal Statistical Society B}, 77:803-825, 2015. 
\vskip 10pt

\hangindent=2em
\hangafter=1
[4] J. Peng, P. Wang, N. Zhou, and J. Zhu. Partial correlation estimation by joint sparse 
regression models. {\it Journal of the American Statistical Association}, 104:735�746, 
2009. 
\vskip 10pt

\hangindent=2em
\hangafter=1
[5] M .Rudelson and R. Vershynin. Hanson-Wright inequality and sub-gaussian 
concentration. {\it Electronic Communications in Probability}, 18:1-9, 2013. 

\end{document}